\documentclass[conference]{IEEEtran}
 
\IEEEoverridecommandlockouts
 \usepackage{cite}
\usepackage{amsmath,amssymb,amsfonts}
\usepackage{algorithmic}
\usepackage{graphicx}
\usepackage{textcomp}
\usepackage{amsmath} 
\usepackage{fourier} 
\usepackage{array}
\usepackage{makecell}
\usepackage{caption}
\usepackage{subcaption}
 \usepackage[T1]{fontenc}        % <-- added
 \usepackage[utf8]{inputenc}     % <-- added
  \usepackage[table]{xcolor}% http://ctan.org/pkg/xcolor
\usepackage{multirow}
\usepackage[colorlinks]{hyperref}
\usepackage{adjustbox}
\usepackage{lipsum}

\usepackage{hhline}

\def\BibTeX{{\rm B\kern-.05em{\sc i\kern-.025em b}\kern-.08em
    T\kern-.1667em\lower.7ex\hbox{E}\kern-.125emX}}
\hypersetup{
	colorlinks=true,       % false: boxed links; true: colored links
	linkcolor=black,        % color of internal links
	citecolor=black,        % color of links to bibliography
	filecolor=magenta,     % color of file links
	urlcolor=black         
}
 \definecolor{light-gray}{gray}{0.9}
 \definecolor{dark-gray}{gray}{0.5}

\usepackage[T1]{fontenc}
\usepackage{array}
\usepackage{multirow}
\makeatletter

\usepackage[english]{babel}

\begin{document}

\title{Network Link Dimensioning based on Statistical Analysis and Modeling of Real Internet Traffic\\ }

%\title{Deploying Dimensioning Network for Real Internet  Traffic: Traffic Properties and Models\\}

\author{\IEEEauthorblockN{Mohammed Alasmar}
\IEEEauthorblockA{\textit{School of Engineering and Informatics} \\
\textit{University of Sussex}\\
Brighton, UK \\
m.alasmar@sussex.ac.uk}
 \and
 \IEEEauthorblockN{Nickolay Zakhleniuk}
 \IEEEauthorblockA{\textit{Computer Science and Electronic Engineering} \\
 \textit{University of Essex}\\
 Colchester, UK \\
 naz@essex.ac.uk}
 
}

 \maketitle

\begin{abstract}
	
Link  dimensioning is used by ISPs to properly provision the capacity of their network links. Operators have to make provisions for sudden traffic bursts and network failures to assure uninterrupted operations. In practice, traffic averages are used to roughly estimate required capacity. More accurate solutions often require traffic statistics easily obtained from packet captures, e.g. variance. Our investigations on real Internet traffic have emphasized that the traffic shows high variations at small aggregation times, which indicates that the traffic is self-similar and has a heavy-tailed characteristics.  Self-similarity and heavy-tailedness are of great importance for network capacity planning purposes. Traffic modeling process should consider all Internet traffic characteristics. Thereby, the quality of service (QoS) of the network would not affected by any mismatching between the real traffic properties and the reference statistical model. This paper proposes a new class of traffic profiles that is better suited for metering bursty Internet traffic streams. 
We employ bandwidth provisioning to determine the lowest required bandwidth capacity level for a network link, such that for a given traffic load, a desired performance target is met. We validate our approach using packet captures from real IP-based networks. 
The proposed link dimensioning approach starts by measuring the statistical parameters of the available traces, and then the degree of fluctuations in the traffic has been measured. This is followed by choosing a proper model to fit the traffic such as lognormal and generalized extreme value distributions. Finally, the optimal capacity for the link can be estimated by deploying the bandwidth provisioning approach. It has been shown that the heavy tailed distributions give more precise values for the link capacity than the Gaussian model.

\end{abstract}


\section{Introduction}
Recently, there is  an increasing demand on high performance services in the Internet; these services include data, voice, and video transmission, these three main services are termed as triple-play services. In the context of IP networks, the validation of the network depends on the examination of the QoS metrics such as delay, delay-jitter, packet loss, availability and throughput \cite{sigcomm20020}. These metrics are described in a committed contract between the users and the service providers which is known as service level agreement (SLA). The above mentioned QoS metrics are mainly relied on bandwidth planning of the network. This indicates the importance of sufficient bandwidth to be provisioned. 

A commonly used bandwidth allocation mechanisms in the IP networks are defined in RFC 1633 and RFC 2475, the two IETF models refer to the integrated services (Intserv) and the differentiated services (Diffserv) respectively \cite{rfc1633,rfc2475}. The complexity of DiffServ and IntServ in deploying the QoS metrics can be avoided by using simple bandwidth provisioning mechanism. The main idea behind bandwidth provisioning is to allocate sufficient bandwidth to the link until achieving satisfied performance, which ensures that the SLA requirements are met\cite{ieee-network-2009}.

In the conventional methods of assurance the link bandwidth, we just apply rules of thumb, such as bandwidth over-provisioning by upgrading the link bandwidth to 30\% of the average traffic value \cite{ieee-network-2009}. This ensures there is no traffic congestion will take place in the link.  The drawback of this mechanism is that it can provide by more bandwidth than is actually needed, intuitively, this will increase the cost of connection. On the other hand, the bandwidth provisioning approach provides the link by the essential bandwidth that guarantees the required performance \cite{computer-networks-2006}. 

\begin{figure}[htbp]
\centerline{\includegraphics[scale=0.13]{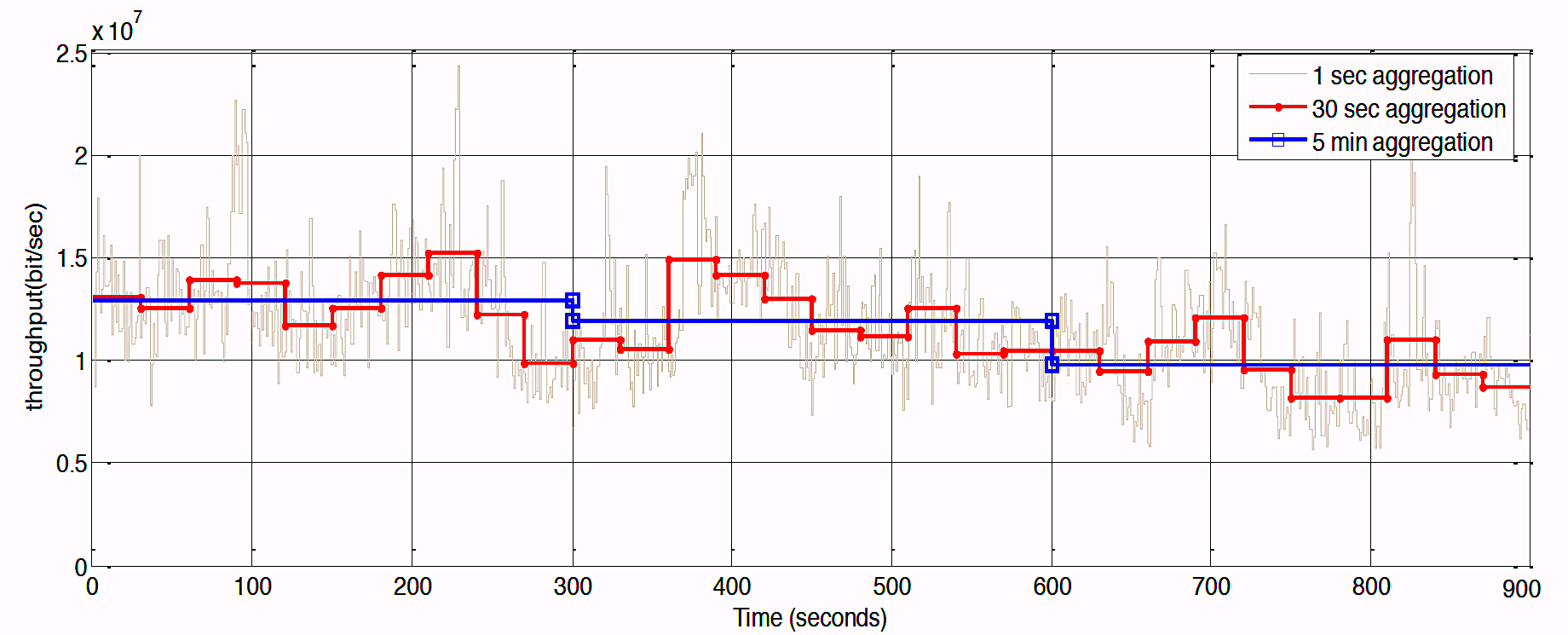}}
\caption{The throughput of a captured traffic at different timescales}
\label{timescales}
\end{figure}

The timescale of traffic aggregation is critical in evaluating the link capacity. Fig. \ref{timescales} shows the throughput (bits/sec) of a captured trance over an interval of 900 sec at different timescales: 1 sec, 30 sec and 300 sec. It is obvious that more fluctuations (burstiness) appear at small values of aggregation time. The more the fluctuations, the more the throughput values are far from the mean, which indicates more variation. Therefore,  the conventional techniques of bandwidth allocation are imprecise at small timescales, and this will break the SLA requirements.    

There is a large body of work aiming to study the Internet traffic properties. Some studies   \cite{ieee-ton-94,ieee-comm-mag,12-GLOBECOM2002,He2006} show that the Internet traffic has the following properties: self-similarity, traffic burstiness and heavy tails. Consequently, the designing of an optimal Internet traffic model should consider these properties. This model plays a critical role in planning networks.

Meent et al. \cite{ieee-network-2009} introduced a new bandwidth provisioning formula, which relies on the statistical parameters of the captured traffic and a performance parameter. They demonstrated that the Internet traffic is bursty over a wide range of aggregation times. They assumed that the Internet traffic can be characterised by  a Gaussian distribution. Their assumption about the applicability of a Gaussian distribution to represent the Internet traffic is based on some related works as  \cite{10-Gaussian-everywhere,9-infocom-2003}. However, this work has missed two major investigations, which are the validation of Gaussianity assumption and the testing of the self-similarity of the Internet traffic. 

In some situations, we cannot observe complete information about the traffic. Employing the efficient estimation method and then choosing the best model in this situation are very important.  In \cite{11-testing}, it is concluded that the tails of the traffic do not track Gaussian distribution, and it is suggested that heavy-tailed distributions are more accurate in representing the Internet traffic.  These distributions have higher peaks and heavier tails than normal distributions. Besides, they have good statistical and reliability properties. Examples of heavy-tailed distributions are Log-normal distribution, Pareto distribution, Weibull distribution, Generalized Extreme Value (GEV) distribution and log-gamma distribution. These distributions are more accurate in representing long-range dependence and self-similar traffic. In this context, the following studies \cite{heavy-sigcomm-2001,14-heavy-tailed-2010-trans,heavy-tails-2013,heavy-tail-sgicom-2007} have reviewed evidence that Internet traffic is characterised by long-tailed distributions.

In this paper we present a statistical analysis and best fitted distribution model of  IP-based Internet traffic. We demonstrate that the Internet traffic is not perfectly fitted with the normal distribution. The fact that Gaussian distribution characterises several aggregated traffics is based on the central limit theorem \cite{2014-ifip-conf}. However, this theory is valid for independent and identically distributed (iid) random processes and it fails if there are dependences between any combinations in the distribution.
Therefore, the resultant empirical performance criterion of bandwidth provisioning approach over Gaussian model does not achieve the target performance, as more attention has to be paid to the tail values. As a result of the fitting tests and the validation of the empirical performance, we found that the lognormal and the GEV models are the proper heavy tails distribution for the network traffic.

This paper makes the following specific contributions.
\begin{itemize}
	\item Investigation self-similarity property in Internet  traffic (see section II), the presence of this property means that the traffic is burstiness which indicates more extreme values present at small aggregation times. 
	\item Finding an optimal statistical model to characterise the Internet  traffic(see section IV), this model has to consider the network traffic's properties. Whilst the traditional network traffic follows a Markovian model, this model is not valid in expressing the Internet  traffic. This failure comes from the fact that the Internet traffic is bursty on a wide range of time scales \cite{ieee-trans-networking}.
	\item Deploying bandwidth provisioning approach over the suggested models (see sections V and VI).\\
\end{itemize}

In our experiments, we used 653 of real network traffic traces that have been captured at five locations , which differ substantially, in terms of both size and the types of users \cite{Meent2010Traces}. The traces were captured over a period of 15 minutes during different times in the day and the night. Table \ref{locations} summarises the details of the five monitored locations at  a university campus.

\begin{table}[ht]

		\arrayrulewidth=0.8pt
	\fontsize{8.5}{8.5}\selectfont
	\setlength\extrarowheight{3pt}
	\centering
	\caption{The details of the monitored locations}
	\label{locations}
	\begin{tabular}{|l|l|l|}
		\hline
	 	\rowcolor[HTML]{EFEFEF} 
		Location                                                                    & Description                                                                                                                                & \begin{tabular}[c]{@{}l@{}}\#Packet \\ traces\end{tabular} \\ \hline
		\begin{tabular}[c]{@{}l@{}}1. Residential \\ network\end{tabular}           & \begin{tabular}[c]{@{}l@{}}300 Mbps Ethernet link connects\\ 2000 students \\ (each has 100 Mbps access link)\end{tabular}                 & 15                                                       \\ \hline
		\begin{tabular}[c]{@{}l@{}}2. Research \\ institute \\ network\end{tabular} & \begin{tabular}[c]{@{}l@{}}1 Gbps Ethernet link connects \\ 200 researchers \\ (each has 100 Mbps access link)\end{tabular}                & 185                                                      \\ \hline
		\begin{tabular}[c]{@{}l@{}}3. Large\\  college\end{tabular}                 & \begin{tabular}[c]{@{}l@{}}1 Gbps Ethernet link connects \\ 1000 employees \\ (each has 100 Mbps access link)\end{tabular}                 & 302                                                      \\ \hline
		\begin{tabular}[c]{@{}l@{}}4. ADSL \\ access\\  network\end{tabular}        & \begin{tabular}[c]{@{}l@{}}1 Gbps ADSL link is used by \\ hundreds of users (each has from \\ 256 kbps to 8 Mbps access link)\end{tabular} & 147                                                      \\ \hline
		\begin{tabular}[c]{@{}l@{}}5. Educational \\ organisation\end{tabular}      & \begin{tabular}[c]{@{}l@{}}100 Mbps Ethernet link connects \\ 135 students and employees\\  (each has 100 Mbps LAN)\end{tabular}           & 4                                                        \\ \hline
	\end{tabular}
\end{table}

\section{Self-similarity in Internet  traffic}

The Internet networks traffic (packet-based networks) performs self-similarity; this is due to the existence of burstiness over a wide range of timescales. In conventional models, the distribution of the packets length of the Internet traffic becomes smoother instead of becoming bursty during the aggregation process. Therefore, the conventional models do not have the ability to represent the Internet networks traffic. Because of the importance of self-similarity phenomenon on modeling real Internet traffic, this section presents the results of the self-similarity tests on the captured traces. Firstly, we discuss the properties of self-similar traffics, which have been concluded at several  studies that are established practically to measure and analysis the statistical characteristics of self-similar traffic of a packet-based networks \cite{ieee-ton-94,ieee-comm-mag,18-Ethernet,self-sim-97-trans}.

\subsection{The correlation function is Long-range dependence (LRD)}

This property measures  the robust dependence between the present values and the old values of any random process.
In order for  any process to be LRD, the autocorrelation function has to decrease hyperbolically, this can be satisfied by getting non-summable autocorrelation function: $\sum\limits_{k=-\infty}^\infty r(k)=\infty$.  In contrast, the autocorrelation function of a short-range dependence processes decrease exponentially \cite{2009-trans-net-LRD}. Hence, for any process that is self-similar, its autocorrelation function can be formulated as follows: 
\[
 r(k) \sim ak^{-\beta} ,  k\rightarrow \infty    \tag{1} \label{equ1}
 \]
where $0< \beta <1$, which is a positive constant value and  $a$ is a scaling factor. This means that the central limit theorem is not applicable on a self-similar traffic, as it is just applicable on iid processes. \\

\subsection{Slowly decaying variance} 
The aggregation process $X_{k}^{(m)}$ can be defined as follows: 

\[
X_{k}^{(m)}=\frac{1}{m}\sum\limits_{i=km-m+1}^{km}X_{i}=\frac{1}{m}(X_{km-m+1}+...+X_{km})      \tag{2} \label{equ2}
 \]

where $X_{k}^{(m)}$  is the averaging value of time series of each non-overlapping neighbouring blocks each of size $m$, while $k$ is the index of each block and  $N$ is the total size of the time series. 
The variance of the aggregated process  $X^{m}$ of a self-similar process can be described as follows \cite{20-self-sim-book}: 

\[
Var(X_{k}^{(m)})=\frac{1}{N/m}\sum_{k=1}^{N/m}(X_{k}^{(m)}-\overline{X^{m}}) \sim  m^{-\beta }   \tag{3} \label{equ3}
 \]
for all  $m=1,2,3,... $ ,   $k=1,2,..., N/m $ and   $0< \beta <1$  where $\overline{X^{m}}$  is the mean value of the segment $X^{(m)} $, which is calculated as follows:

\[ 
 \overline{X^{m}}=\frac{1}{N/m}\sum_{k=1}^{N/m}X_{k}^{(m)}
 \]

\subsection{The power spectrum has a power-law distribution around zero frequency } 
For any discrete sequence $X_{n}$ of length N, the power spectrum $ S(\omega)$ can be calculated by using the discrete Fourier transform (DFT) or the fast Fourier transform (FFT) as follows \cite{20-self-sim-book}:

\[
 S(\omega)=\frac{1}{2\pi N}\left | \sum_{n=1}^{N}X_{n} e^{jn\omega } \right |^{2}  \tag{4} \label{equ4}
 \]
 where $ \omega=2\pi n/N $ and  $-\pi \leq \omega \leq \pi $

 The power spectrum of a self-similar process follows a power law and it is centred at the origin. This can be described as follows:
 \[
 S(\omega)\sim \left | \omega \right |^{-\delta }   \textrm{    as  } \omega\rightarrow 0  ,\textrm{       } 0<\delta <1   \tag{5} \label{equ5}
 \]

 \section{The statistical tests of Self-similarity }
There are several statistical tests to examine the self-similarity of any distribution or process. We are going to use three of these tests: R/S test, time-variance plot, and Periodogram method. The estimated value of the Hurst parameter in each test gives evidence about the presence of the self-similarity in the examined trace. If $ 0.5<H<1 $ this implies that the traffic is self-similar \cite{20-self-sim-book}.\\

\subsection{Variance-Time Test }
The logarithm of the two sides in equation (3) gives:
  \[
 log\left [ var(X_{k}^{(m)}) \right ] \sim -\beta log(m) , \textrm{ as}\ m\rightarrow \infty    \tag{6} \label{equ6}
 \]

  The value of $\beta$  is obtained from the log-log plot: $  log \left [ var(X_{k}^{(m)}) \right ]  $  versus  $log(m)$   where   $\beta$  is the slope of the fitted line (using least-squares method) in this plot. For slowly decaying variance the slope of the fitted line $ \beta$ has to be between  -1 and 0. Where the Hurst parameter is equal to  $H=1- \beta /2$.\\

 \subsection{Rescaled-Range (R/S) Test}
 R/S test  \cite{1997-self-sim} demonstrates to measure the variability in the traffic. For a series  $X(t)= X_{1}, X_{2}, ... , X_{N}$  of $N$ samples and sample mean $\mu _{n}$. The range R is the difference between the maximum and minimum values of the cumulative summation of the deviation values at each sample point, as follows: 
 
  \[
R(n)= \underset{1<t<n}{max}  \sum_{i=1}^{t}(X_{i}-\mu _{n})-\underset{1<t<n}{min}  \sum_{i=1}^{t}(X_{i}-\mu _{n})  \tag{7} \label{equ7}
 \]
 The relation between Hurst parameter and R/S value can be described as follows:  
 \[
 E\left [ \frac{R(n)}{S(n)} \right ]  \sim cn^{H},  \textrm{as}\  n\rightarrow \infty   
\]
where  S(n) is the standard deviation of the samples and $c$ is a positive constant number. The logarithm of the two sides gives:
 \[
 log\left(   E\left [ \frac{R(n)}{S(n)} \right ]    \right)   \sim H log(n)+log(c)    \tag{8} \label{equ8}
 \]
 Thus, the Hurst parameter is equal to the slope of the fitting line on the log-log graph:  $ log\left(   E\left [ \frac{R(n)}{S(n)} \right ]    \right) $ versus $log(n)$.\\

 \subsection{Periodogram Test}

\begin{figure}[ht]
\centerline{\includegraphics[scale=0.11]{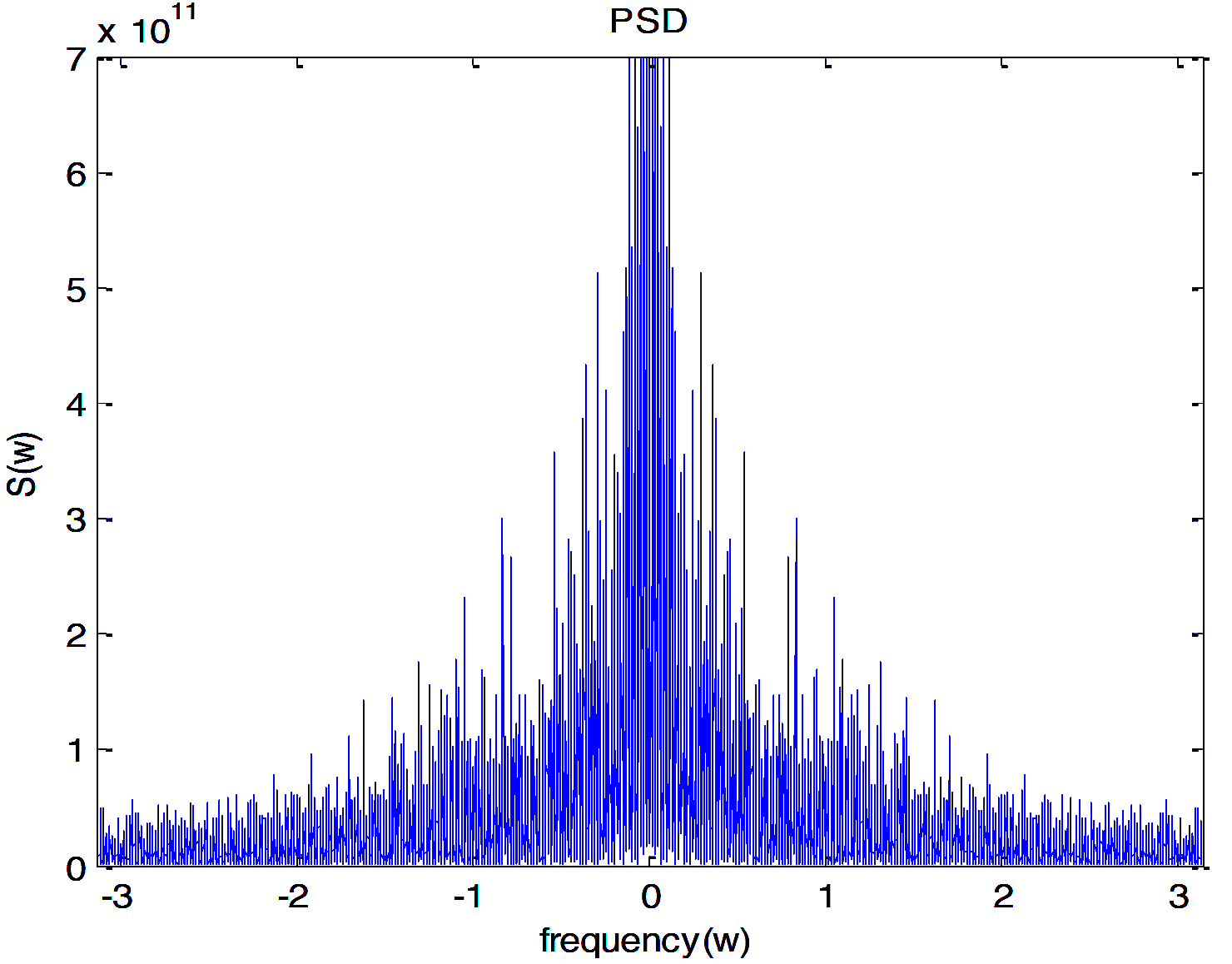}}
\caption{The power spectrum density of an Internet traffic}
\label{fig2powerSpectrum}
\end{figure}

This method is characterised as a frequency domain estimation of the Hurst parameter, and it is more precise than the previous tests. The advantage of this method over the previous tests comes from the fact that there is no need for aggregation or combination of the original traffic points during the calculation of the power spectrum.  The power spectrum density $ S(\omega )$ can be calculated from equation (4). It is important to note that the existence of self-similarity will affect the power spectrum at the band of the low frequencies i.e. as $\omega\rightarrow 0$. This indicates that the power spectrum of self-similar process follows a power law distribution as $\omega\rightarrow 0$, as shown in Fig. \ref{fig2powerSpectrum}.

From equation (5)  $ S(\omega )$  can be obtained as follows:  
 \[
 S(\omega ) \sim \left | \omega \right |^{1-2H}  \textrm{  ,  as}\ \omega\rightarrow 0  \tag{9} \label{equ9}
 \]
 
\begin{figure}[ht]
\begin{center}
\begin{subfigure}[b]{0.3\textwidth}
\centering
\includegraphics[scale=0.11]{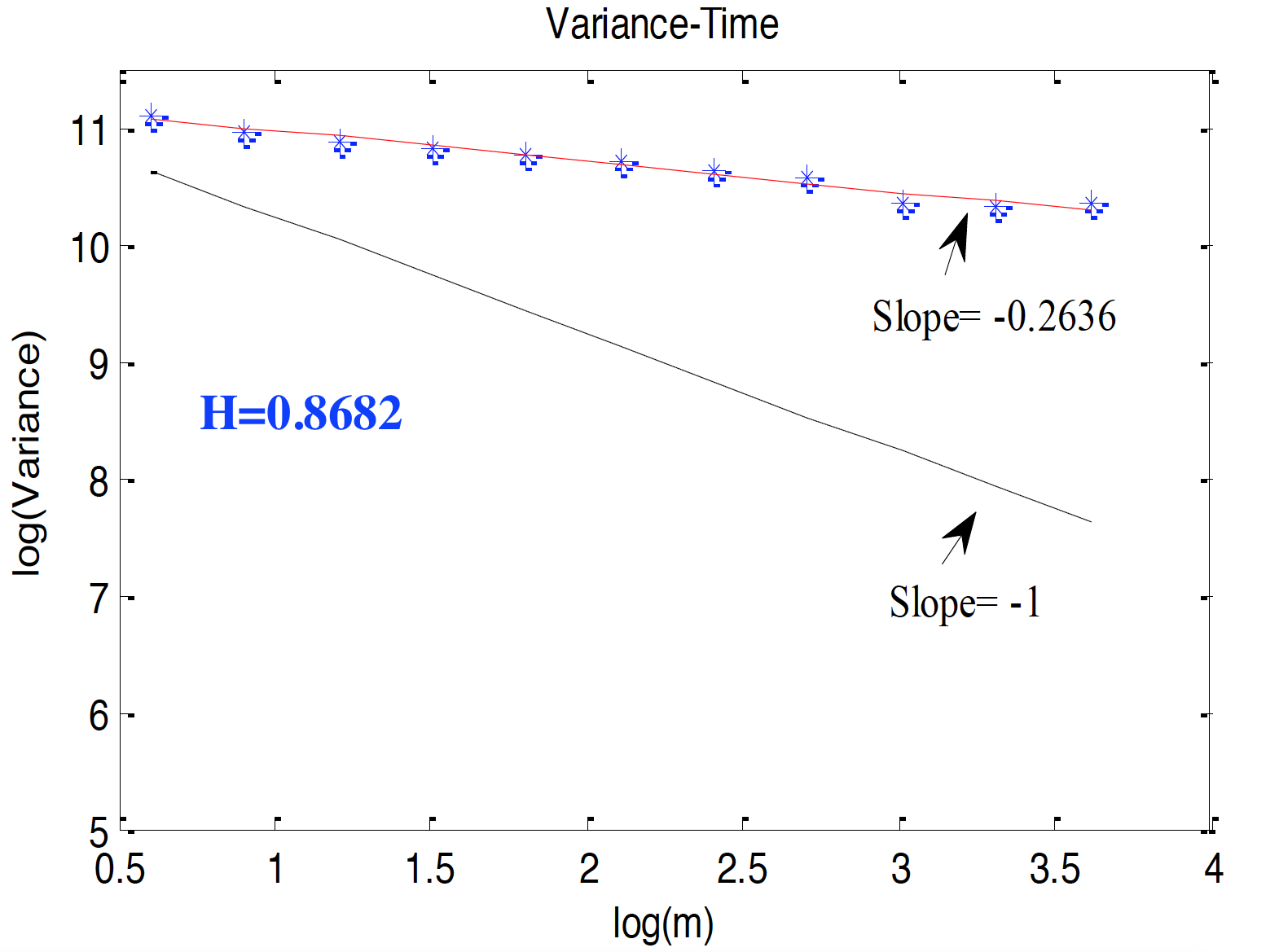}
\caption{V-T}
\hspace*{0.2em}
\end{subfigure}
\begin{subfigure}[b]{0.3\textwidth}
\centering
\includegraphics[scale=0.11]{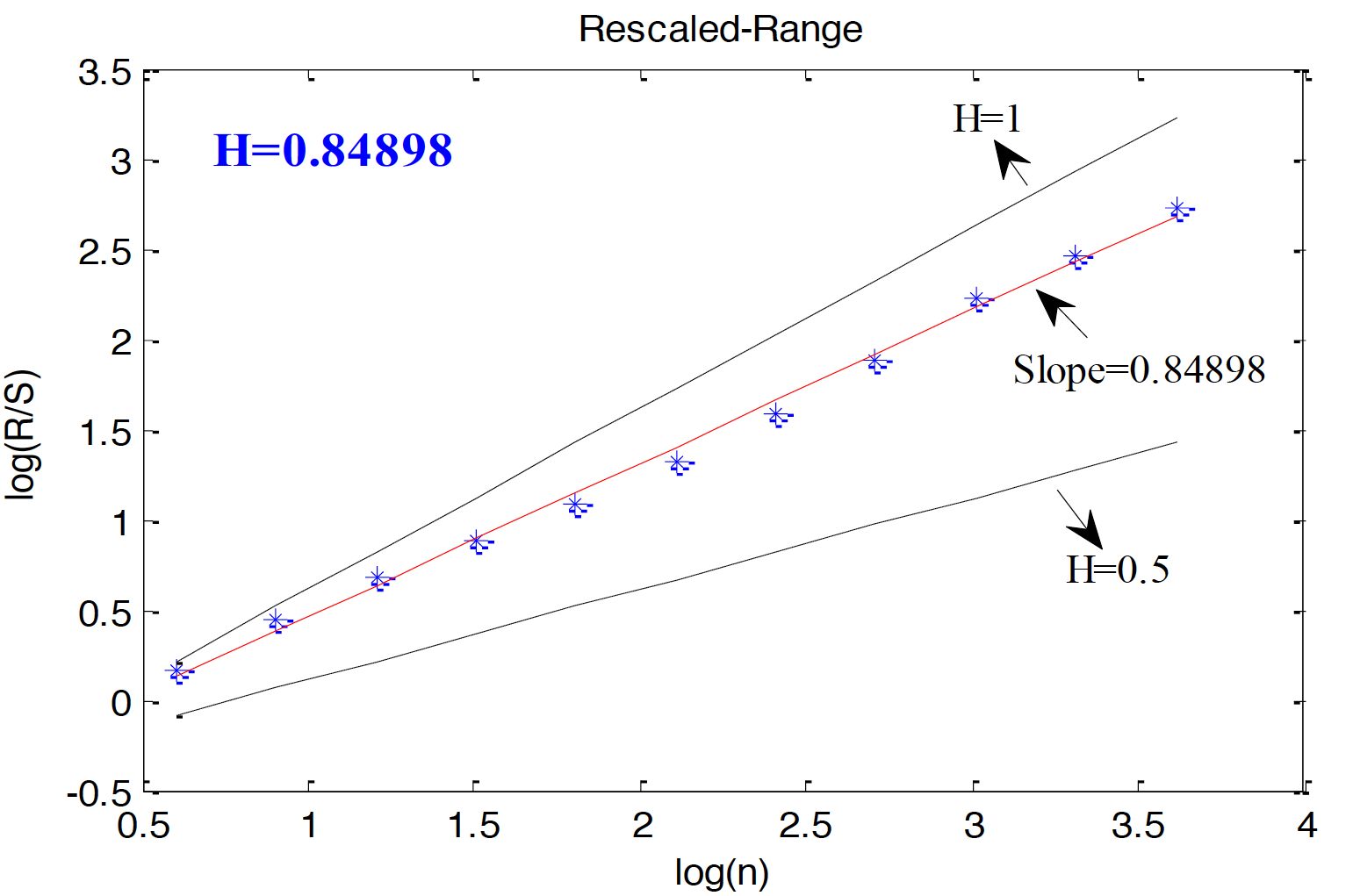}
\caption{R/S}
\hspace*{0.2em}
\end{subfigure}
\begin{subfigure}[b]{0.3\textwidth}
\centering
\includegraphics[scale=0.11]{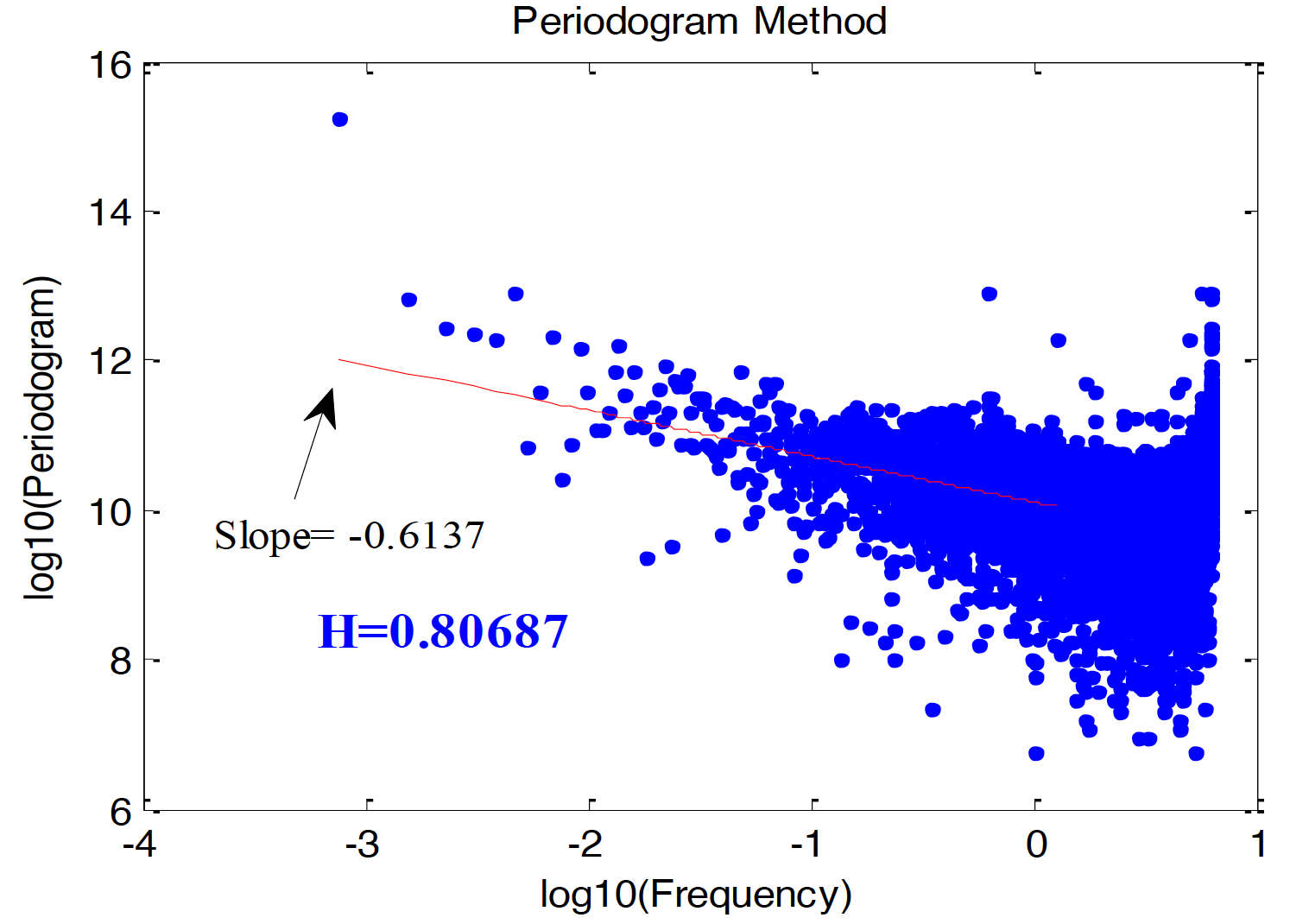}
\caption{Periodogram}
\end{subfigure}
\caption{Plots of graphical estimators of Trace1}
\label{selfsimtests} 
\end{center}
\end{figure}

The logarithm of the axes in Fig. \ref{fig2powerSpectrum} produces the log-log plot of which is shown in Fig. \ref{selfsimtests}(c).
The estimation of H is done by fitting only 10\% to 20\% of the lower frequencies, which is our region of interest, as the behaviour of the power spectrum function describes in equation (9) is not applicable for high frequencies values. The slope of the straight fitted line is related to the Hurst parameter as follows: $H=(1-slop)/2$.\\

\textbf{Results of self-similarity tests.}
The testing of self-similarity is based on estimating the Hurst exponent value (as discussed in section III). Fig. \ref{selfsimtests} shows the produced figures from the self-similarity tests on one of the captured traces. As depicted in Fig. \ref{selfsimtests}, the value of the Hurst exponent from different tests is close to 1, which implies that the tested trace is self-similar.

We experimentally validate that the IP traffic is self-similar by running the same tests on hundreds of traces. Table \ref{hurst} displays the results of the self-similarity tests on 6 different traces from different locations.

\begin{table}[ht]
	
	\arrayrulewidth=0.5pt
\fontsize{8.5}{8.5}\selectfont
\setlength\extrarowheight{5pt}
	\centering
	\caption{Estimation Hurst parameter using V-T, R/S and Periodogram tests}
	\label{hurst}
	\begin{tabular}{|c|c|c|c|c|}
		\hline
		\rowcolor[HTML]{EFEFEF} 
		\cellcolor[HTML]{EFEFEF}{\color[HTML]{000000} } & \cellcolor[HTML]{EFEFEF}{\color[HTML]{000000} } & \multicolumn{3}{c|}{\cellcolor[HTML]{EFEFEF}{\color[HTML]{000000} Estimated value of H}  } \\ \hline
		\rowcolor[HTML]{EFEFEF} 
		\multirow{-2}{*}{\cellcolor[HTML]{EFEFEF}{\color[HTML]{000000}  \begin{tabular}[c]{@{}c@{}}Traces \\  \end{tabular} }} &
		
		 \multirow{-2}{*}{\cellcolor[HTML]{EFEFEF}{\color[HTML]{000000} \begin{tabular}[c]{@{}c@{}}Number \\ of packets\end{tabular}}} & {\color[HTML]{000000}  V-T test  } & {\color[HTML]{000000} R/S test} & {\color[HTML]{000000} Peri. test} \\ \hline
		\rowcolor[HTML]{FFFFFF} 
		1 & 399,326 & 0.9087 & 0.8283 & 0.8040 \\ \hline
		\rowcolor[HTML]{FFFFFF} 
		2 & 3,110,382 & 0.8789 & 0.7637 & 0.8069 \\ \hline
		\rowcolor[HTML]{FFFFFF} 
		3 & 3,747,745 & 0.8691 & 0.8302 & 0.9102 \\ \hline
		\rowcolor[HTML]{FFFFFF} 
		4 & 5,864,815 & 0.8679 & 0.7699 & 0.8789 \\ \hline
		\rowcolor[HTML]{FFFFFF} 
		5 & 4,304,742 & 0.8333 & 0.8594 & 0.9534 \\ \hline
		\rowcolor[HTML]{FFFFFF} 
		6 & 4,213,812 & 0.8962 & 0.8062 & 0.9233 \\ \hline
	\end{tabular}
\end{table}

Again the results show that the Internet traffic is self-similar, as  H has a value close to 1. 

We developed a Matlab GUI tool that is aimed to run the same tests over any available traces \cite{ github , YouTube  }. More traces are available here \cite{Meent2010Traces}. 

From this section, it is concluded that the Internet traffic is self-similar. This indicates that the traffic is bursty for wide range of timescales. 
\\

 \section{Best fitted distribution model of internet traffic}

In this section, we investigate whether the traffic in our traces is accurately described by a Gaussian process or by other heavy-tailed distributions. We employ the following probability distributions matching tests: quantile-quantile plot, probability density functions (PDFs) matching plot and correlation coefficient test.

\subsection{The quantile-quantile plot (Q-Q plot)}
  Q-Q plot is a powerful visualization test which can assess the degree of similarity between different distributions. The main idea of the Q-Q plot is to compare the observed data with one of the well-known distributions such as normal distribution. The quintiles are defined as the values that are taken from the random variables every regular interval. The x-axis of the Q-Q plot represents the quantiles of the reference distribution while y-axis represents the quantiles of the observed samples \cite{transaction2015}. Q-Q plots are created by plotting the pair
  
  \[
  \left ( F^{-1}\left ( \frac{i}{n+1} \right ),S(i) \right ),\textrm{     } i=1,... ,n  \tag{10} \label{equ10}
  \]
 where n is the number of samples, $F^{-1}$  is the inverse cumulative distribution function of the reference distribution and $S(i)$  is the observed samples. The two distributions are matched if the scattered points of the two quantiles follow a straight fitting line.

\subsection{The linear correlation coefficient test}
The covariance measures the strength of the relation between two random variables. For strong measuring of goodness-of-fit, the normalized version of the covariance, which is known as linear correlation coefficient can be used. The normalization factor is the multiplication of the standard deviation of both: the empirical distribution standard deviation $\sigma_{S_{(i)} } $ and the reference distribution standard deviation $\sigma_{x_{i} }$. Hence, the correlation coefficient can be written as \cite{transaction2015}: 
  
     \[
  \gamma =\frac{cov(S_{(i)},x_{i})}{\sigma_{S_{(i)} }\sigma_{x_{i} }}= \frac{\sum_{i=1}^{n} \left ( S_{(i)}- \hat{\mu} \right )\left ( x_{i}-\bar{x} \right )}{\sqrt{\sum_{i=1}^{n}\left ( S_{(i)}-\hat{\mu} \right )^{2}.\sum_{i=1}^{n}\left ( x_{i}-\bar{x} \right )^{2}}}  \tag{11} \label{equ11}
    \]
  
where $S_{(i)}$ is the observed samples, and its mean value:   $\hat{\mu}=\frac{1}{n}\sum_{i=1}^{n}S_{(i)} $, while $x_{i}$ is the reference distribution samples which can be calculated from the inverse CDF of the reference random variable: $x_{i} =F^{-1}\left ( \frac{i}{n+1} \right )$  and its mean value: $\bar{x}=\frac{1}{n}\sum_{i=1}^{n}x_{i} $.
  
The value of the correlation coefficient can vary between:  $-1\leq \gamma \leq 1$. Note that as the value of $\gamma$ changes from $\pm 1$  to 0, then relation strength will drop from strong to moderate and finally to weak strength around the zero value. 
For the purpose of getting stronger goodness-of-fit, the acceptable value of $\gamma$ is suggested to be above 0.95.\\
  
\textbf{Results of testing the matching between the Internet  traffic and the suggested models } 

\begin{figure}[htp]

\begin{center}

\begin{subfigure}[b]{0.4\textwidth}
\centering
\includegraphics[scale=0.15]{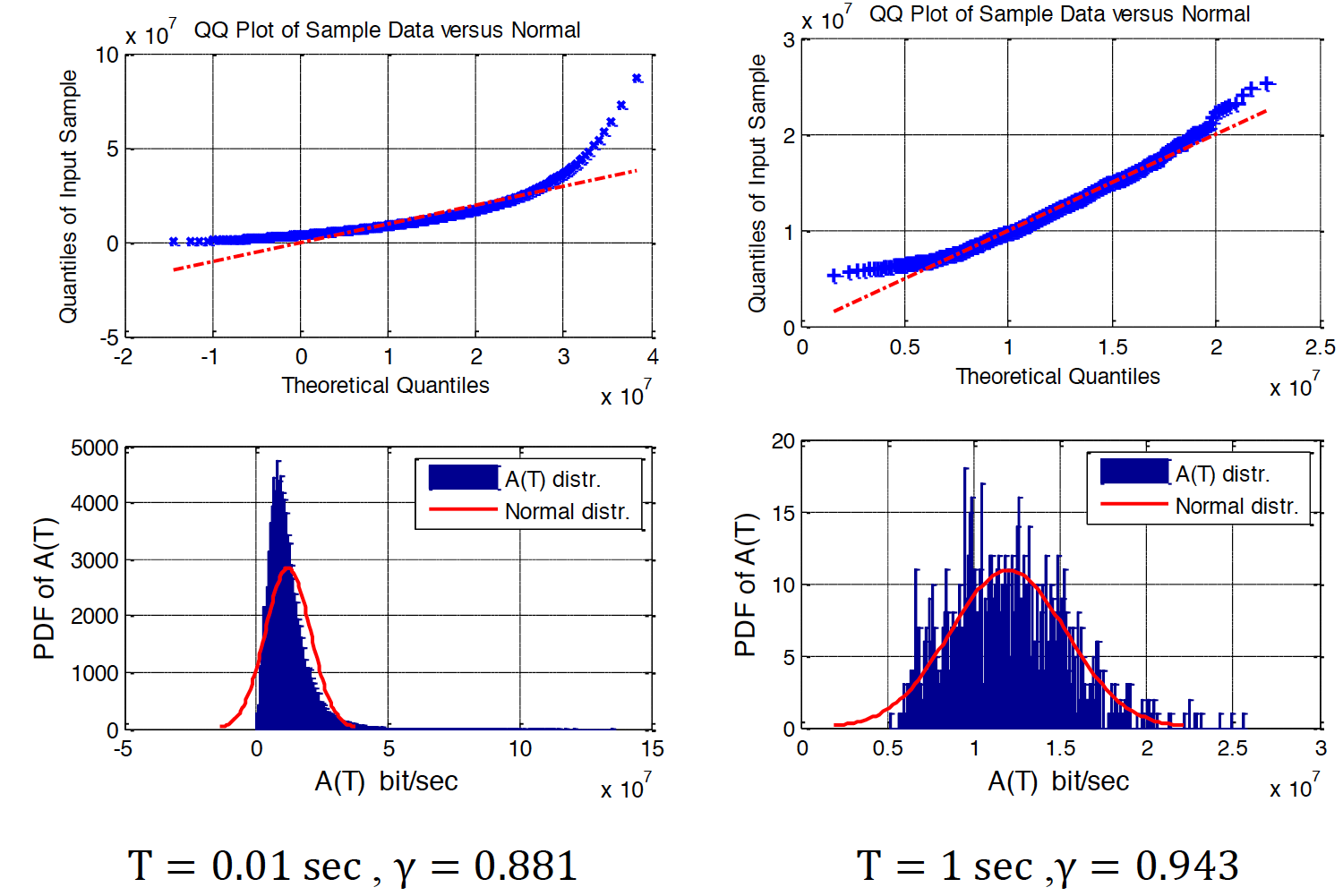}
\caption{  }
\end{subfigure}
\begin{subfigure}[b]{0.4\textwidth}
\centering
\includegraphics[scale=0.15]{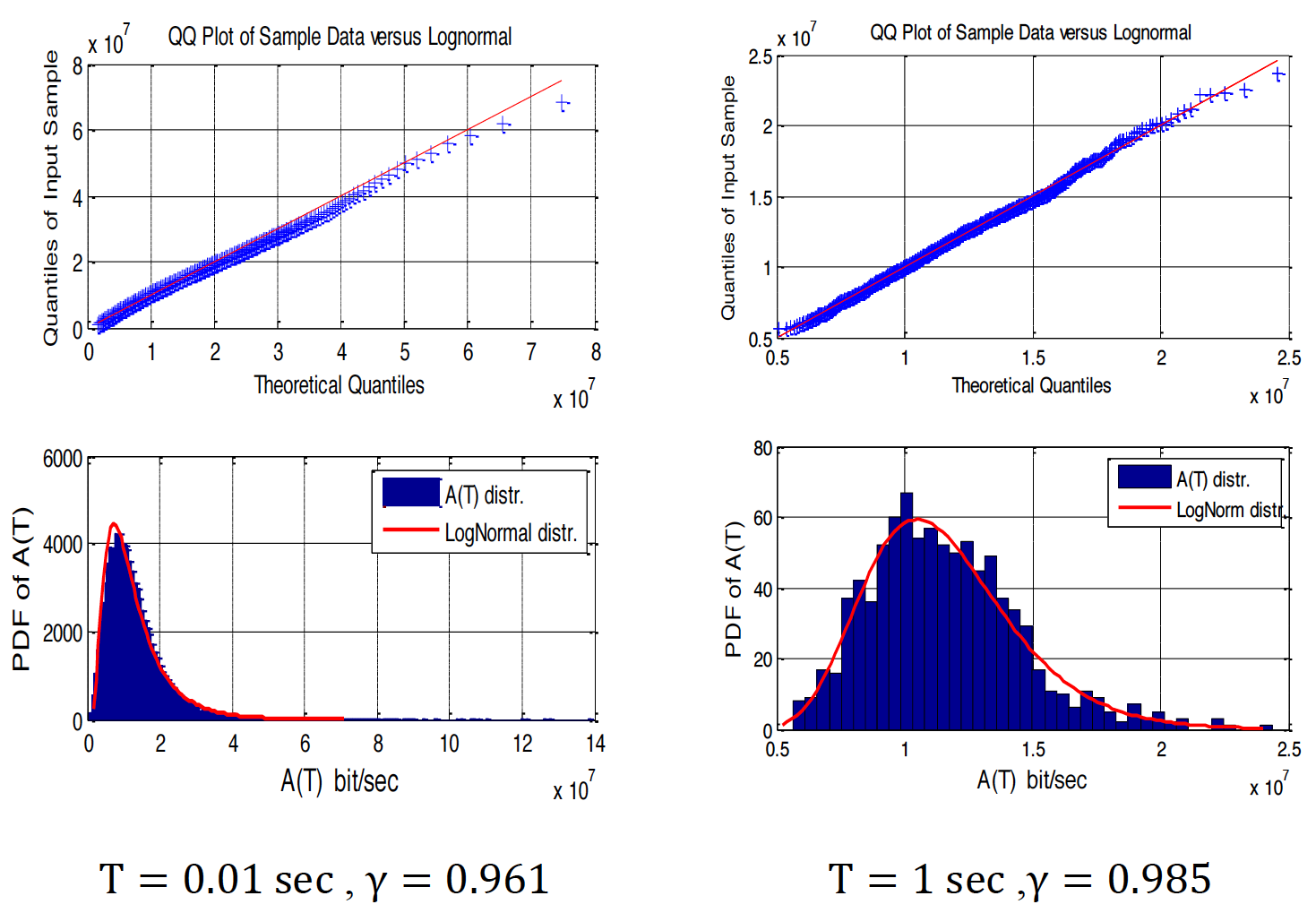}
\caption{  }

\end{subfigure}

\begin{subfigure}[b]{0.4\textwidth}
\centering
\includegraphics[scale=0.15]{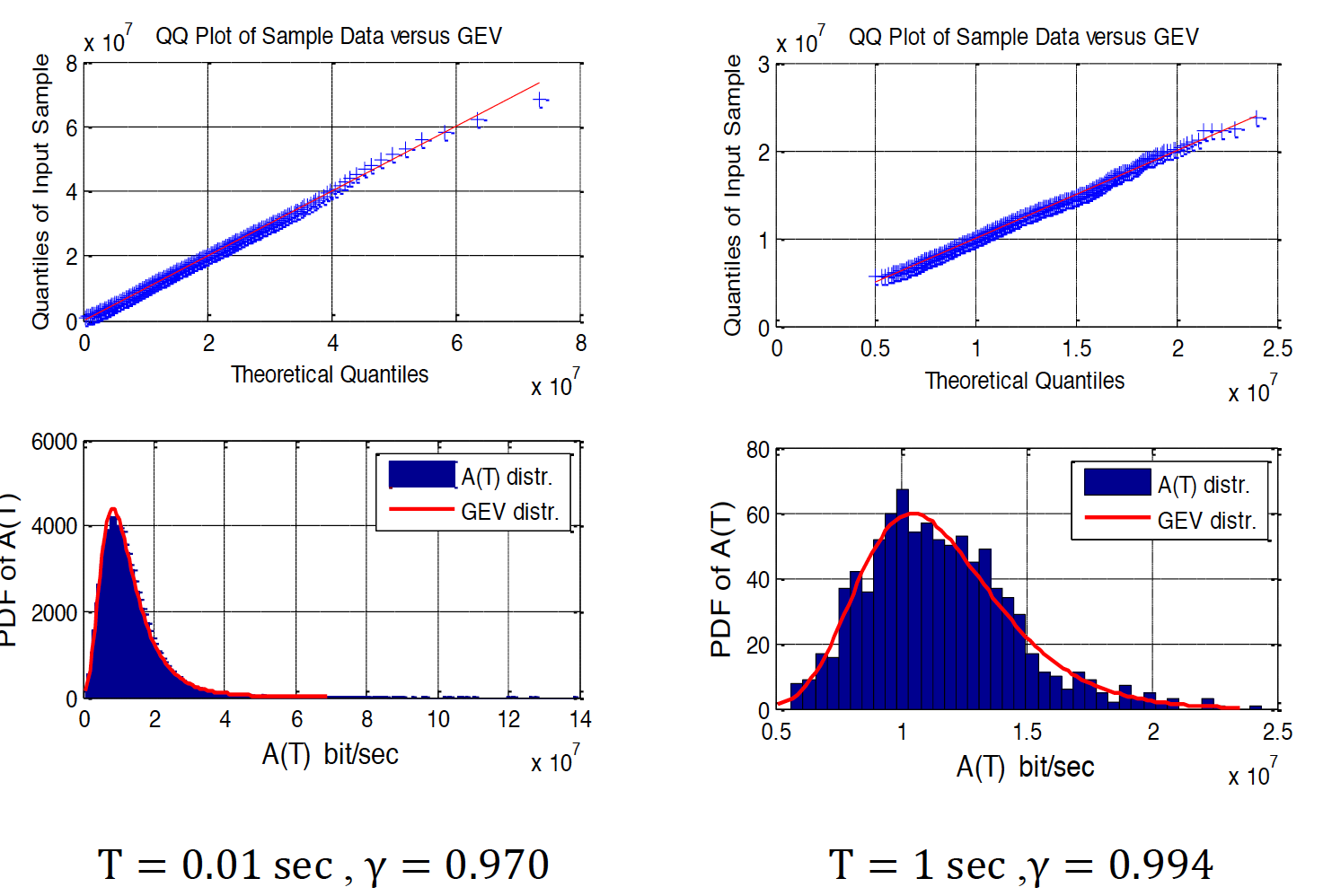}
\caption{ }
\end{subfigure}

\caption{Q-Q plot, value and PDF at different timescales for (a) Normal (b) Lognormal (c) GEV distributions }
\label{pdfqqqq} 
\end{center}
\end{figure}

Fig. \ref{pdfqqqq} shows the results of applying Q-Q plot test on Trace1 at different aggregation times (T=0.01 sec and T=1 sec) and by using different reference distributions (Normal, Lognormal and GEV).  Besides, it shows the PDF of both the captured traffic (the blue bars in the sub-figures) and the fitting curve of the reference distributions (the red curves in the sub-figures). The aggregation times are chosen to be reasonably small to include traffic fluctuations as discussed in the introduction.

From Fig. \ref{pdfqqqq}(a), it is noticeable that the captured traffic is not perfectly fitted with the normal distribution. The mismatching between the two distributions takes place at the tails values, as the Q-Q points are deviated from the straight lines at the tails. In addition, the correlation coefficient values do not indicate strong correlation between the distributions, as the values of  $\gamma$ are below 0.95. Based on the analysis of these results, it is obvious that we need to pay more attention to the tails. On the other hand, Fig. \ref{pdfqqqq}(b)-(c) show that the captured traffic is perfectly fitted with the GEV distribution and it is almost fitted to the lognormal distribution. In addition, the correlation coefficient values indicate strong correlation between the distributions, since the values of $\gamma$  are larger than 0.95. Unlike normal distribution, the tails in the Q-Q plots show some extreme values which could not be represented by the reference lines.

 It is concluded that heavy-tailed distributions such as lognormal and GEV distributions are more accurate in representing the Internet  traffic. This conclusion is based on testing hundreds of traces from different locations by using our developed tool \cite{github}. The obtained results are always close to the results shown in Fig. \ref{pdfqqqq}.

\section{Bandwidth provisioning approach }

In this section we deploy the bandwidth provisioning mechanisms based on the statistic distributions that have been proved to be more accurate in characterising the Internet traffic. The goal of the bandwidth provisioning approach is to enhance the channel capacity without extra bandwidth. Although the normal model is not suitable to represent the network traffic as explained in the previous section, we will continue under the gaussianity assumption to demonstrate that this model gives unsatisfactory results in deploying the bandwidth provisioning approach. 

In the bandwidth provisioning approaches, link transparency is selected as QoS criteria. The following inequality is used for the purpose of accurate checking of the link transparency \cite{transaction2015}:\\
  \[
  P\left ( A(T)\geq CT \right ) \leq \varepsilon   \tag{12} \label{equ12}
  \]
  
It is obvious from this inequality that the probability of finding the captured traffic over a specific period of timescale  $A(T)/T$  larger than the link capacity has to be smaller than the value of the performance criterion $\varepsilon$. The value of $\varepsilon$ represents the probability of packet loss and its value has to be chosen carefully by the network provider in order to meet the specified SLA; in general, $\varepsilon$ has to be below the probability $10^{-2}$. Likewise, the value of the aggregation time T should be sufficiently small so that the fluctuations in the traffic can be modeled as well.

Mainly, optimising the link capacity will be investigated based on the following five bandwidth provisioning (BWP) approaches:\\
- \textbf{Approach 1:} BWP through direct calculations under the standard normal PDF \\
- \textbf{Approach 2:} BWP using extended formula of the Gaussian model   \\
- \textbf{Approach 3:} BWP using Meent's approximation formula  \\
- \textbf{Approach 4:} BWP based on  lognormal distribution model  \\
- \textbf{Approach 5:} BWP  based on GEV distribution  model   \\

\subsection{Approach 1: BWP through direct calculations under the standard normal PDF}
Under the Gaussianity behaviour, the captured traffic $A(T)$ is described as follows, 

\[
A(T) \sim Norm(\mu T,\upsilon (T)) \tag{13} \label{equ13}
\]

where  $\mu_{A}=\mu T$ is the mean value (in bits) and $\upsilon (T)$ is the variance  (in bits$^{2}$) of $A(T)$.

\begin{figure}[htbp]
\centerline{\includegraphics[scale=0.14]{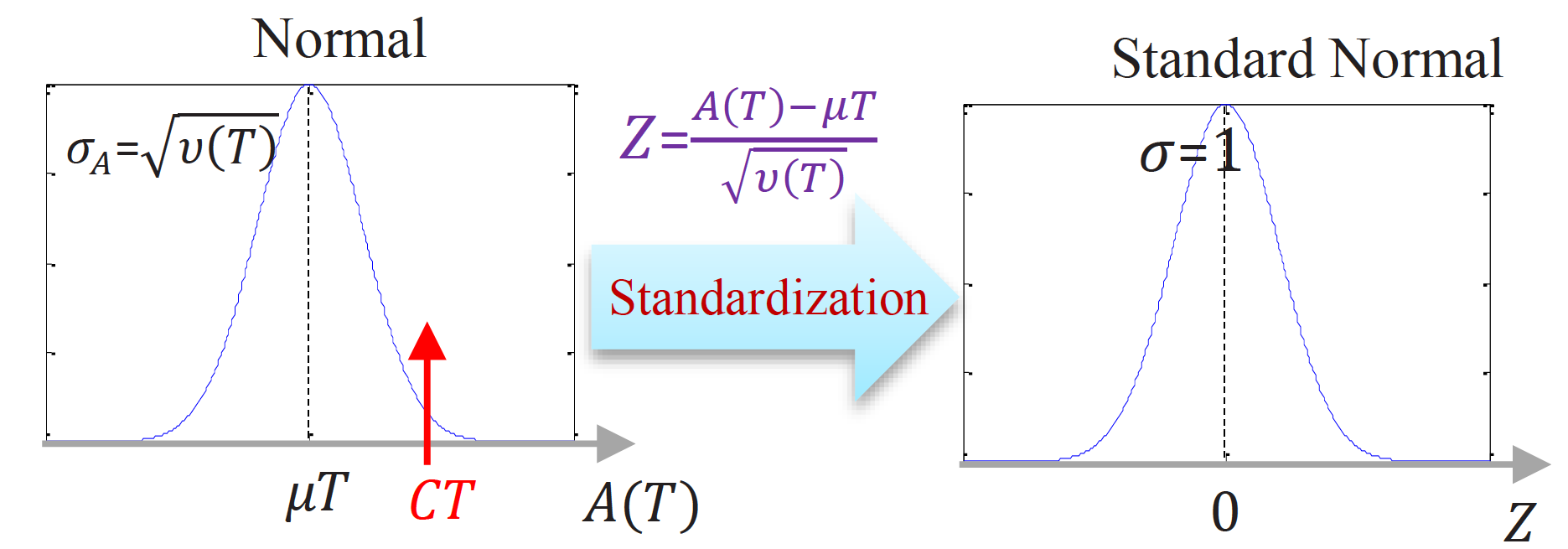}}
\caption{The normal and the standard normal distribution of A(T)}
\label{gaussTran}
\end{figure}

The link transparency condition (equation 12) can be solved by finding the value of C which satisfies the value of the performance criterion $\varepsilon$, as shown in Fig. \ref{gaussTran}. 
For more simplicity, the variable $A(T)$ has to be standardized by mapping it from normal distribution $A(T)$ to standard normal distribution $Z$. This will make the probabilities calculations simpler. This transformation is given by: 

\[
A(T) =\mu T+\sqrt{\upsilon (T)} Z   \tag{14} \label{equ14}
\]

By substituting for $A(T)$ from equation (14) into equation (12) we obtain:

  \[
P\left ( \mu T+\sqrt{\upsilon (T)} Z \geq CT \right )    \leq \varepsilon  
  \]

Then,

   \[
P\left (  Z\geq \frac{CT-\mu T}{\sqrt{\upsilon (T)}}\right ) \leq \varepsilon   \\
  \]

Now,

  \[
T\left (  \frac{CT-\mu T}{\sqrt{\upsilon (T)}}\right )=1-\Phi \left (  \frac{CT-\mu T}{\sqrt{\upsilon (T)}} \right )\leq \varepsilon    \tag{15} \label{equ15}
 \]

where $T(z)$ is the complementary cumulative distribution function (CCDF) of the standard normal distribution, and $\Phi(z)$ is the cumulative function, where  $T(z)=1-\Phi (z) $ and $ \Phi (z) =P\left ( Z\leq z \right ) $.

Hence, from equation (15) the value of the link capacity can be written as follows: 

 \[
  \textrm{\textbf{C1:}}\  C= \Phi ^{-1}\left ( 1-\varepsilon  \right )\sqrt{\frac{\upsilon (T)}{T^{2}}}+\mu    \tag{16} \label{equ16}
\]
where $C$ is the link capacity (in bits per second), $\upsilon (T)$ is the variance of the captured traffic (in bits$^{2}$), $\mu $ is the mean traffic rate (in bits per second) and T is the aggregation time.
It is noticeable from equation (16) that the required capacity increases by decreasing the value of the aggregation time T. In addition, the lower values of  $\varepsilon$ indicate more capacity is needed.

  \subsection{Approach 2: Bandwidth provisioning using extended formula of the Gaussian model}
  
 In the case of Gaussianity assumption of the network traffic, the transparency formula can be solved by finding the area under the tails of the Gaussian PDF. The tail function is defined as:
   
\[
 T(z)= P(Z>z) \approx  \frac{1}{z\sqrt{2\pi }}e^{-\frac{1}{2}z^{2}} \textrm{ ,    as } z\rightarrow \infty   \tag{17} \label{equ17}
\]
      
Equation (15) can be approximated to:
    
\[   
T\left (Z\geq   \frac{CT-\mu T}{\sqrt{\upsilon (T)}}\right )= \frac{1}{ \left (  \frac{CT-\mu T}{\sqrt{\upsilon (T)}}\right ) \sqrt{2\pi }} e^{-  \frac{1}{2}\left (\frac{CT-\mu T}{\sqrt{\upsilon (T)}}\right )^{2}} \leq \varepsilon 
 \]
 
 Hence,
 
  \[ 
  \textrm{\textbf{C2:}}\ \frac{ \left (CT-\mu T  \right )^{2}}{\upsilon (T)}+log \left (\frac{2\pi \left (CT-\mu T  \right )^{2}}{\mu T}  \right )\geq -2log(\varepsilon ) \tag{18} \label{equ18}
  \]
       
Thus, in the second approach the link capacity can be evaluated by solving equation (18).

  \subsection{Approach 3: Bandwidth provisioning using Meent's formula } 

As discussed in the Introductory section, Meent et al. \cite{ieee-network-2009} suggested a new formula to deploy bandwidth provisioning mechanism. Although this study has not mentioned the tail presence in the real network traffic, the suggested formula has been proved by initiation of tail bounds inequalities. It is necessary to bound the tails values, these tails refer to the random processes which deviate far from its mean. Meent et al. used Chernoff bound where tails are represented exponentially, as follows:

\[ 
P  \left  ( A(T)\geq CT  \right )\leq   \frac{E\left [e^{SA(T)}  \right ]}{e^{SCT}}=e^{-SCT}E\left [e^{SA(T)}  \right] \tag{19} \label{equ19}
\]

where ${E\left [e^{SA(T)} \right ]}$ is the moment generation function (MGF) of the captured traffic $A(T)$.

Meent's dimensioning formula to find the minimum link capacity is defined as follows \cite{ieee-network-2009}:

\[
\textrm{\textbf{C3:}}\  C=\mu +\frac{1}{T} \sqrt{-2log(\varepsilon) .\upsilon (T)}  \tag{20} \label{equ20}
\]
    
From equation \ref{equ20}, the link capacity is obtained by adding safety margin value to the average of the captured traffic,
    
\[
    \textrm{Safety margin = }\sqrt{-2log(\varepsilon)}  \textrm{ . } \sqrt{\frac{\upsilon (T)}{T^{2}} }
 \]
 
 This safety margin value depends on the performance criterion $\varepsilon$  and the ratio $\sqrt{\upsilon (T)/T^{2}}$. As the value of  $\varepsilon$ decreases the safety margin will increase. For example, the value of the safety margin increases by 40\% as the value of  $\varepsilon$ decreases from $10^{-2}$ to $10^{-4}$. 
 
\begin{figure}[htbp]
\centerline{\includegraphics[scale=0.14]{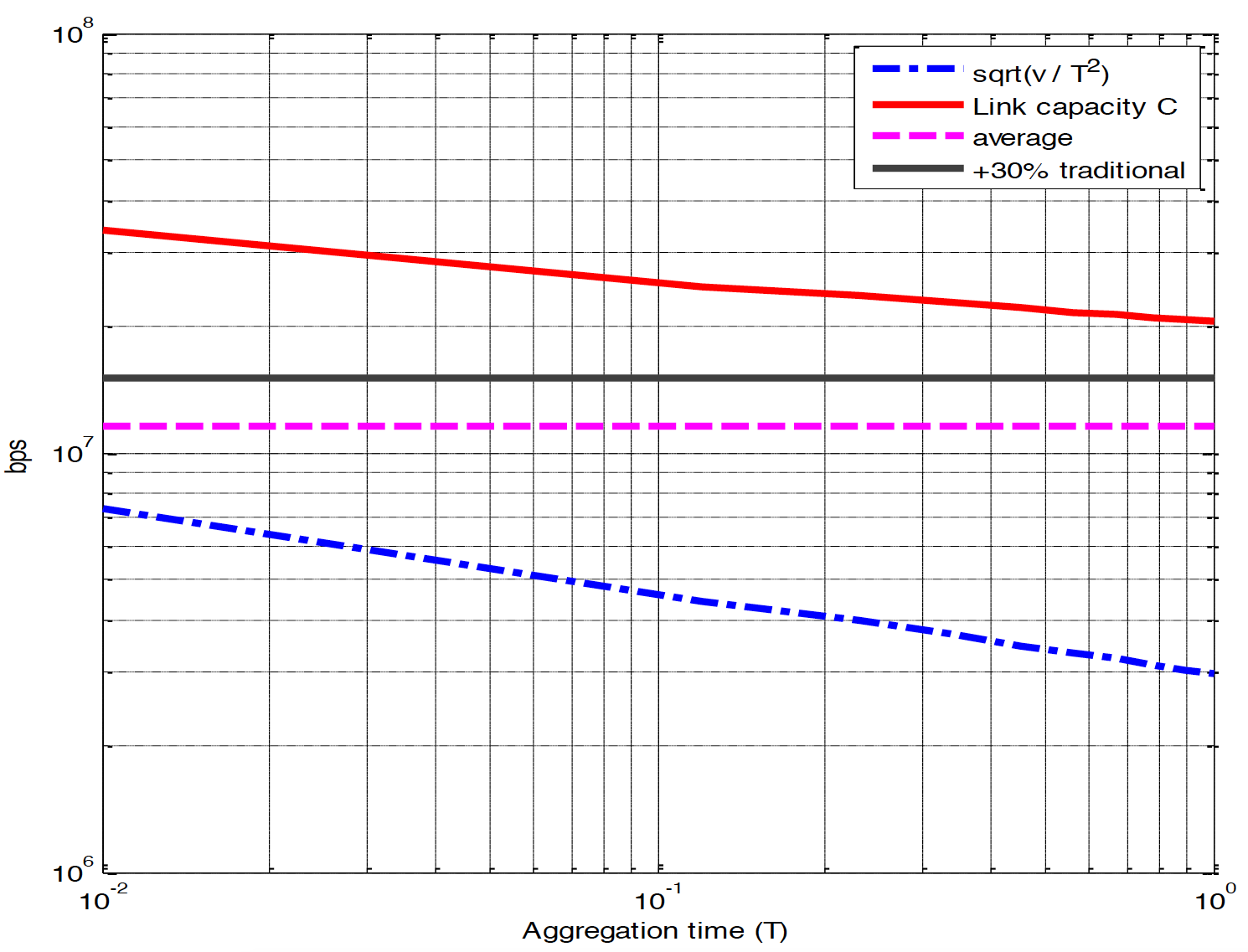}}
\caption{Comparison between bandwidth provisioning approach and the traditional approach}
\label{30trad}
\end{figure}

Fig. \ref{30trad} shows that the link capacity formula is in line with the notion of bandwidth provisioning, as the deployed bandwidth on the link is changed with the variation of the traffic characteristics: $\mu$ and $\upsilon (T)$. This is different from the conventional techniques, where the safety margin is fixed to be 30\% above the average of the presented traffic.

 \subsection*{\textbf{Empirical value of Network layer loss}}

  Practically, for TCP/IP network and as specified in RFC2680
the acceptable packet loss rate should be below 1\% \cite{rfc2680}. Moreover, IEPM group at Stanford Linear Accelerator Center (SLAC) reported that the percentage of packet loss between 0-1\% indicates a good network performance, while the percentage between 1-2.5\% can be acceptable \cite{loss-stanford}. In addition, it has been reported by Pingman \cite{pigman-tools} (which is a company that builds software that makes network troubleshooting suck less) that packet loss larger than 2\%  over a period of time is a strong indicator of problems.

\begin{figure}[ht]
	\centerline{\includegraphics[scale=0.245]{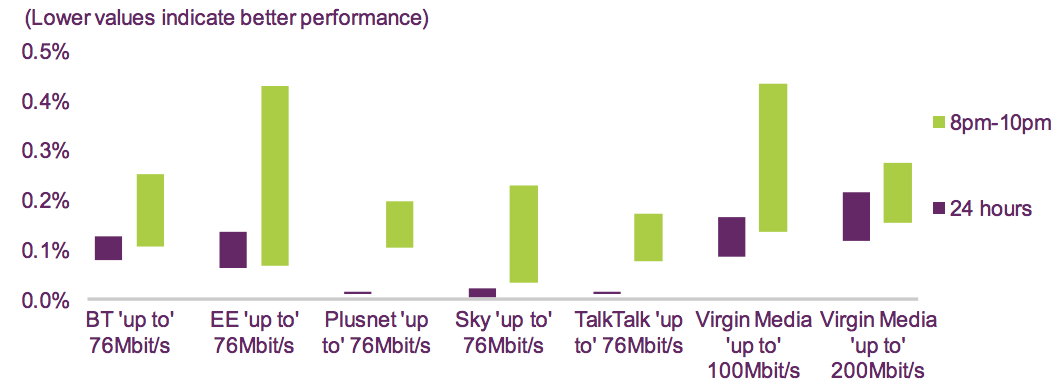}}
	\caption{Ofcom packet loss report on November 2016 from different  ISP panel members }
	\label{ofcom}
\end{figure}
Furthermore, Ofcom the UK’s communications regulator \cite{ofcom2016} has reported the average and peak-time packet loss for ISP packages on November 2016 from different  ISP panel members: BT `up to' 76Mbit/s, EE `up to' 76Mbit/s, Plusnet `up to' 76Mbit/s, Sky `up to' 38Mbit/s, TalkTalk `up to' 38Mbit/s, Virgin `up to' 100Mbit/s and Virgin `up to' 200Mbit/s, as shown in Fig. \ref{ofcom}.

Lately, the NTT Europe Ltd (which is rated as one
of the top ranked telecommunication companies in the world) has reported 
in one of its modified SLA report \cite{ntt-eu} that packet loss rate  has to be 0.1\% or less for Intra-Europe Network and
0.3\% or less for the other NTT Backbone networks.
 \\ 
 \subsection*{\textbf{The validation of the model}}
The validation condition refers to the empirical value of
performance criterion, which is denoted by $\hat{ \varepsilon}$, and it is given by:

 \[
  \hat{ \varepsilon}= \frac{\# \left \{ A_{i}| A_{i}>CT\right \}}{n}  \textrm{ , } i\in 1...n  \tag{21} \label{equ21}
  \]

 This empirical value is defined as the percentage of all the
points of the captured traffic which excess the estimated link
capacity. It has to be less than the target value of the
performance criterion $\varepsilon$, i.e. $  \hat{ \varepsilon}\leq \varepsilon $.
The difference between both values $\hat{ \varepsilon}$ and $\varepsilon$ is due to the fact
that the chosen model is not suitable to characterise the real network traffic.

 \subsection*{\textbf{Comparison between the three Gaussianity assumption approaches}}

Table \ref{threeApproaches} summarises the above discussed bandwidth provisioning approaches. The table includes equations (16), (18) and (20).

 \begin{table}[ht]
	\arrayrulewidth=0.7pt
	\fontsize{9.5}{9.5}\selectfont
	\setlength\extrarowheight{14pt}
	\centering
	
	\caption{The three approaches of bandwidth provisioning based on Gaussian distribution model}
	
	\label{threeApproaches}

	\begin{tabular}{|l|l|}
		\hline
		\rowcolor[HTML]{EFEFEF} 
		Bandwidth provisioning approaches                                                                                                                                           &  Tail representation                              \\ \hline
		\setlength\extrarowheight{38pt}
		$\textbf{C1 } :C= \Phi ^{-1}\left ( 1-\varepsilon  \right )\sqrt{\frac{\upsilon (T)}{T^{2}}}+\mu     $                                                                                & $T(z)_{1} = 1-\Phi(z) $                                                               \\ \hline
		$	\textbf{C2} :  \makecell{  \frac{ \left (CT-\mu T  \right )^{2}}{\upsilon (T)}+log \left (\frac{2\pi \left (CT-\mu T  \right )^{2}}{\mu T}  \right ) \\ \geq -2log(\varepsilon ) }$ & $\makecell{ T(z)_{2}\approx  \\ \frac{1}{z\sqrt{2\pi }}e^{-\frac{1}{2}z^{2}}    }$    \\ \hline
		$\textbf{C3}: C = \sqrt{-2log(\varepsilon)}  \textrm{ . } \sqrt{\frac{\upsilon (T)}{T^{2}} } +\mu $                                                                                                 & $  \makecell{ P \left ( X\geq z  \right ) \\   \leq e^{-SX}E\left [e^{SX}  \right] }$ \\ \hline
	\end{tabular}
\end{table}

\begin{figure}[h]
	\centerline{\includegraphics[scale=0.19]{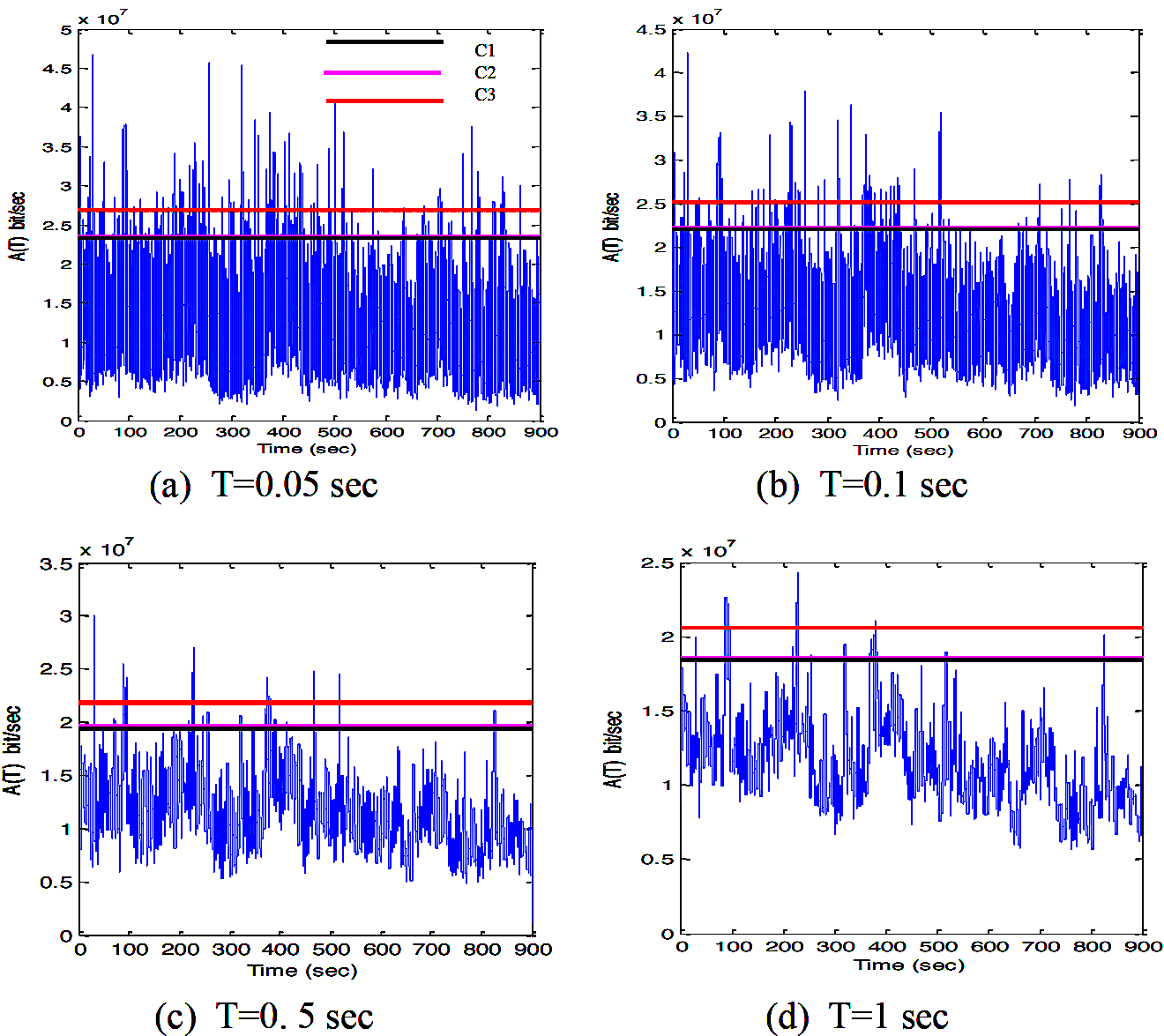}}
	\caption{The captured traffic A(T) and the estimated link capacities at different timescales}
	\label{ATc1c2c3}
\end{figure}

 Fig. \ref{ATc1c2c3} shows the plotting of the captured
traffic $A(T)$ (Trace1) over 15 minutes at different timescales (T= 0.05, 0.1, 0.5 and 1 sec). The three approaches (lines C1,C2,C3 in Fig. \ref{ATc1c2c3}) give different link capacities, which do not match the minimum required capacity, that is expected as these approaches do not characterise the tails accurately. 

Table \ref{123approaches} shows the results of employing the bandwidth provisioning formulas of the first three approaches on trace1 at different timescales.

\begin{table}[ht]
	
	\caption{Bandwidth provisioning results based on Gaussian distribution model}
	\label{123approaches}
	
\setlength\extrarowheight{5pt}
\fontsize{7.5}{7.5}\selectfont

	\centering
	\begin{tabular}{|c|c|c|c|c|c|c|c|}
		\hline
		\rowcolor{dark-gray}
		\multicolumn{8}{|c|}{Target $ \varepsilon  = 0.01$ , mean: $\mu=11.56  Mbps$}                                                                                                                                                                                                                                                                                                                               \\ \hline
		
		\rowcolor{light-gray} 
		&   & \multicolumn{2}{c|}{ \begin{tabular}{c} Approach 1    \end{tabular}}     
		& \multicolumn{2}{c|}{  \begin{tabular}{c} Approach 2\end{tabular}}   
 	& \multicolumn{2}{c|}{  \begin{tabular}{c} Approach 3 \end{tabular}}                                                                                               
		\\  \hline 
		
		\rowcolor{light-gray} 
		\multirow{-2}{*}{\begin{tabular}{@{}c@{}}T \\ (sec)\end{tabular}} 
		&   \multirow{-2}{*}{\begin{tabular}{@{}c@{}}$\upsilon (T)$  \\ Tbps$^{2}$\\  \end{tabular}}
		
		&   \begin{tabular}[c]{@{}c@{}}  C1 \\ Mbps\end{tabular} &   \begin{tabular}[c]{@{}c@{}}Emp.\\ $\hat{ \varepsilon}$\end{tabular} &  
	     	\begin{tabular}[c]{@{}c@{}}C2 \\ Mbps\end{tabular} &  \begin{tabular}[c]{@{}c@{}}Emp.\\ $\hat{ \varepsilon}$\end{tabular} &
	     	\begin{tabular}[c]{@{}c@{}}C3 \\ Mbps\end{tabular} &  \begin{tabular}[c]{@{}c@{}}Emp.\\ $\hat{ \varepsilon}$\end{tabular}

	     	\\ \hline
0.01                                                                                       & 54.4                                                                                             & 28.72                                              & 0.0293                                             & 29.08                                              & 0.0278                                            & 33.94                                              & 0.0135                                            \\ \hline
0.05                                                                                       & 25.6                                                                                               & 23.33                                              & 0.0262                                             & 23.58                                              & 0.0244                                            & 26.92                                              & 0.0120                                            \\ \hline
0.1                                                                                        & 20.4                                                                                               & 22.07                                              & 0.0248                                             & 22.89                                              & 0.0228                                            & 25.27                                              & 0.0111                                            \\ \hline
0.5                                                                                        & 11.6                                                                                               & 19.46                                              & 0.0250                                             & 19.63                                              & 0.0238                                            & 21.87                                              & 0.0177                                            \\ \hline
1                                                                                          & 88.8                                                                                               & 18.49                                              & 0.0288                                             & 18.64                                              & 0.0277                                            & 20.60                                              & 0.0188                                            \\ \hline

	\end{tabular}
\end{table}

Intuitively, there are three questions to ask about the obtained results. Firstly, why do the three approaches give different
values for the link capacity, although all the approaches
are based on the Gaussian distribution model?
Simply, the answer is that in the three approaches the tails are
represented in different approximation, see Table \ref{threeApproaches}.

Secondly, why do not the empirical values $\hat{ \varepsilon}$  satisfy 
the goal performance (Target $ \varepsilon  = 0.01$)?
The answer is established in section IV, where the Gaussian
model is considered as a weak model to represent heavy-tailed Internet traffic.

Thirdly, why does the third approach give best empirical results
in comparison with the first two approaches? This can be inferred from the value of the empirical performance criterion $\hat{ \varepsilon}$ in Table \ref{123approaches}, which is almost around the target value 0.01 at approach 3. In order to answer this
question, it is required to examine the accuracy of representing the tails among the three approaches.
As illustrated in Table \ref{123approaches}, the nearest model to characterise the heavy tails is the third approach, where the tails are
bounded exponentially. In contrast, in the first two approaches the tails are modeled approximately based on the Gaussian model, which is not fitted for heavy-tailed distributions.

\begin{figure}[h]
	\centerline{\includegraphics[scale=0.13]{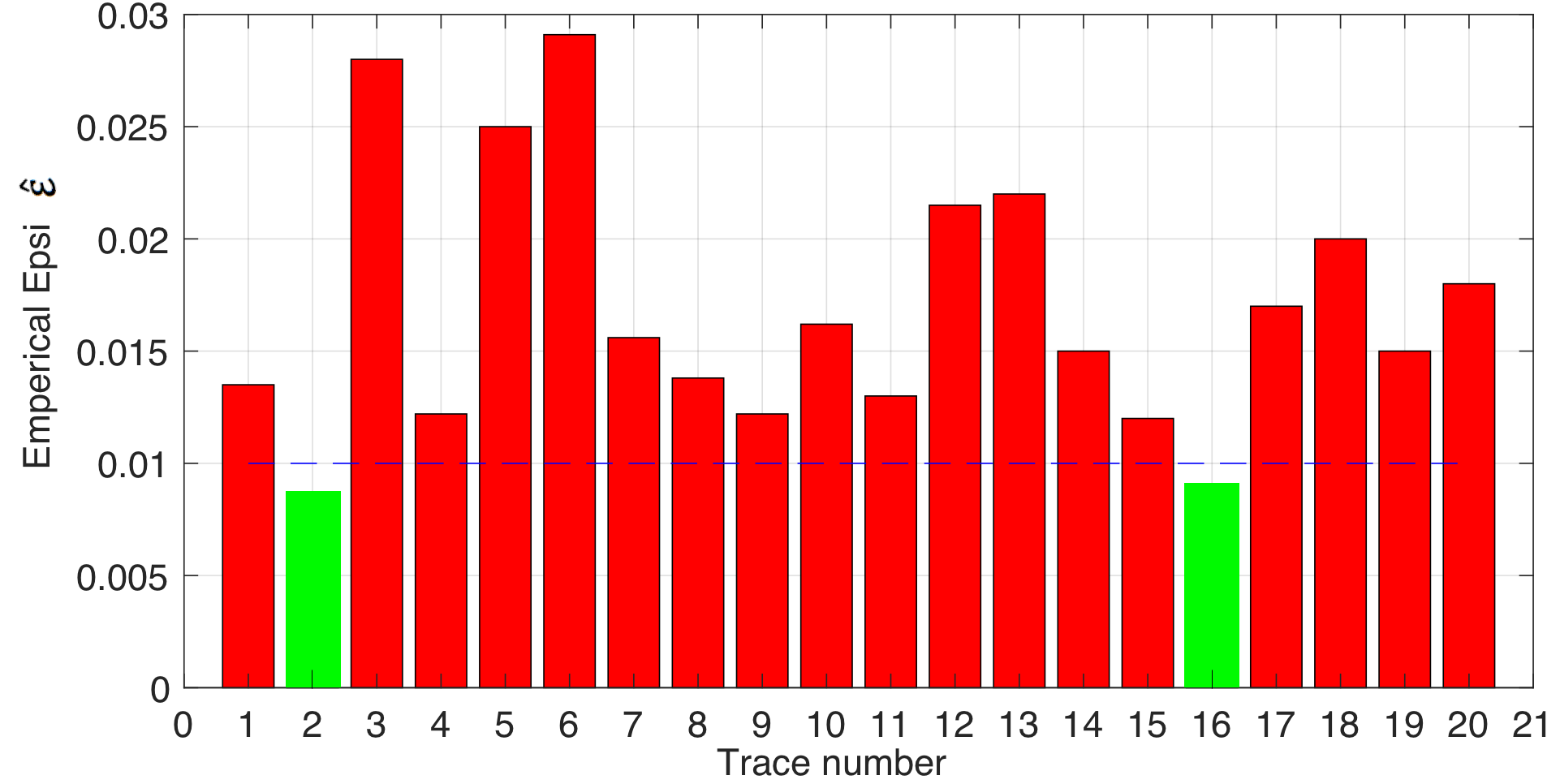}}
	\caption{The empirical performance criterion $\hat{ \varepsilon}$ of 20 traces, when $T=0.01 sec$ and $\varepsilon  =0.01$}
	\label{emp-app123}
\end{figure}

Figure \ref{emp-app123} shows the results of the above described experiment over 20 traces from different locations. The aggregation time of all the captured 20 traces is $T=0.01 sec$ and the performance criterion is $\varepsilon  =0.01$. As expected, most of the traces do not achieve the targeted performance; 18 traces have $\hat{ \varepsilon}$ values larger than 0.01, and just  trace 2 and trace 16 get acceptable link capacities.

The last two bandwidth provisioning approaches (Approach 4 \& Approach 5) are discussed in the next section.  

\section{TRAFFIC MODELING USING HEAVY-TAILED DISTRIBUTIONS   }

\subsection{Heavy tails}
Heavy or fat tails processes are the processes which have
plenty of values far from the mean value. As explained in Fig. \ref{timescales}, the network traffic is bursty at small aggregation times; this
causes the presence of heavy tail in the distribution of the
network traffic. The burstiness in the traffic considers as the
main source of network traffic self-similarity.
The random variable A is said to be distributed with `Heavy Tails'  if \cite{heavytails2017}:

\[
P(A>x)\sim x^{-\alpha }   \textrm{ ,    as } x\rightarrow \infty \textrm{   ,  } 0<\alpha <2   \tag{22} \label{equ22}
 \]

 This indicates that the distribution above large value x of the
random variable A is decreasing hyperbolically instead of
exponentially. For example, the Gaussian distribution does not
consider as heavy-tailed model, because it has exponentially
bounded tails (see equation 17).

Alternatively, the lognormal, Weibull, Pareto
and generalized extreme value (GEV) distributions are good
models for heavy-tailed distributions.

\begin{figure}[ht]
\centerline{\includegraphics[scale=0.20]{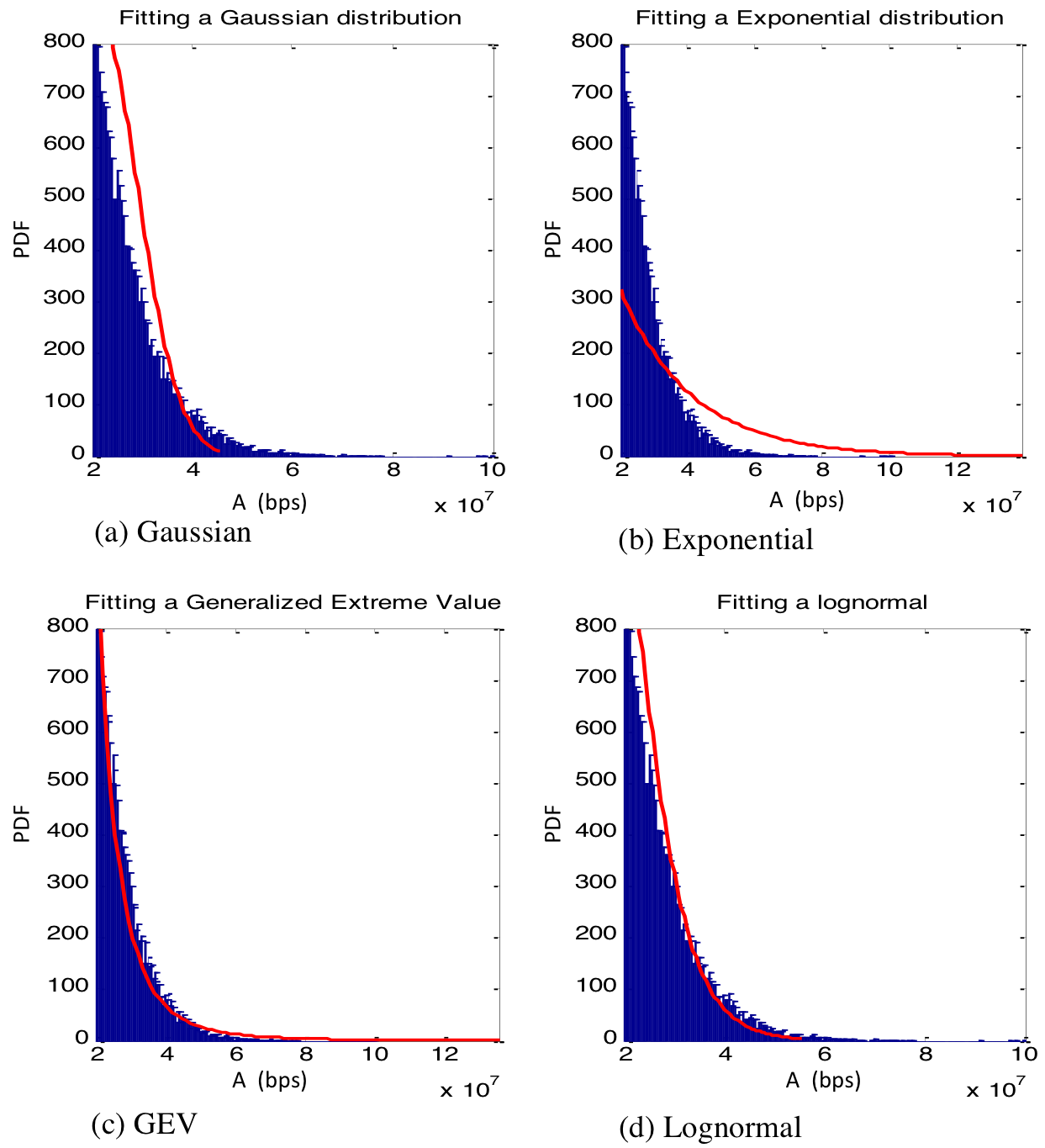}}
\caption{The reliability of different models in representing the tails}
\label{tailspdf}
\end{figure}

Fig. \ref{tailspdf} shows the results of representing the Internet traffic tails values using four different distributions. Evidently,  GEV and lognormal distributions are more accurate than normal and exponential distributions in bounding the tails. Therefore,  lognormal and  GEV models will be chosen as proper heavy tails distribution for the network traffic.

 \subsection{Bandwidth provisioning based on lognormal (Approach 4) and GEV (Approach 5)}
 
The  Internet traffic is better modeled by lognormal and GEV distributions. We will investigative whether these models can satisfy the target performance when employing bandwidth provisioning mechanism. The calculations of the link capacity in these approaches can be done directly through the PDF or CDF functions of the proposed distributions. The following steps describe the bandwidth provisioning approach 4 and approach 5: 
\begin{itemize}
	\item Measuring the statistics parameters: mean ($\mu$)  and variance ($\sigma^2$) of the captured traffic
	\item Specifying the performance criteria $\varepsilon$ that provides the required SLA
	\item Generating a lognormal or GEV distribution (PDF or CDF) based on the measured
	statistics parameters of the captured traffic
	
	\item Applying the link transparency formula (equation 12)
	\item The link capacity can be found by calculating the inverse value of the CDF function at  $1 -\varepsilon$
	
\end{itemize}

Now, we apply the above mentioned steps on  the captured Trace1, which has the mean $\mu = 11.556 Mbits/sec$ and the variance   $\sigma^2 = 2.0412 \times 10^{13} bits^{2}$, when $T = 0.1$ sec.  The performance criteria $\varepsilon$ is chosen to be 0.01. Theses parameters are used to generated a lognormal distribution.

The PDF of a lognormal random variable A(T) is defined as follows:

\[
f(A(T))= \frac{1}{A(T)\sqrt{2\pi \sigma }}e^{-\frac{\left ( log(A(T))-\mu  \right )^{2}}{2\sigma ^{2}}   } \textrm{   ,       } A(T)>0
\]
Thus, the link transparency formula (equation 12) can
be written as follows:

\[
P(A(T)\geq CT)=\int_{A=CT}^{\infty } \frac{1}{A(T)\sqrt{2\pi\sigma}}e^{-\frac{[ log(A(T))-\mu ]^{2} }{2\sigma^{2}}} dA \leq \varepsilon    \tag{23} \label{equ23}
\]

The CDF function of the lognormal distribution that characterises the captured  traffic A(T) is defined as: $ F(C)=P(A(T)/T <C)  $, hence 
\[
 F(C) = P\left (\frac{A(T)}{T} <  C  \right )\geq  1-\varepsilon     \tag{24} \label{equ24}         
\]
This implies that the probability of getting the captured traffic
$A(T)/T$ less than the channel capacity has to be above 0.99.

Finally, the link capacity can be found by finding the inverse of
the lognormal CDF function, as follows:

\[
C=F^{-1}\left ( 1-\varepsilon \right )   \tag{26} \label{equ26}  
\]
Hence, for the captured Trace1 the link capacity can be found as:
$C=F^{-1}\left ( 0.99 \right )= 26.7314 Mbps.$

Bandwidth provisioning based on GEV distribution follows the same above mentioned steps, where the used CDF in equation (26) has to be a GEV distribution CDF function. 

Table \ref{c4c5table} shows the results of the last two approaches: C4 and
C5. In addition, it presents the empirical values of the
performance criteria $\hat{ \varepsilon}$. All the results are measured from
Trace1. The calculated values of $\hat{ \varepsilon}$ indicate that the target
performance of the transmitted packets through the link has
been achieved from both distributions, as all the measured $  \hat{ \varepsilon}$ are less than $\varepsilon $. The results from both approaches are acceptable and better than the first three approaches.

\begin{table}[ht]

\caption{Bandwidth provisioning results based on lognormal and GEV distributions}
\label{c4c5table}

 \arrayrulewidth=0.7pt
\setlength\extrarowheight{9.5pt}
\fontsize{10}{10}\selectfont
\centering
\begin{tabular}{|c|c|c|c|c|c|}
\hline
\rowcolor{dark-gray}
\multicolumn{6}{|c|}{Target $ \varepsilon  = 0.01$ , mean: $\mu=11.56  Mbps$}                                                                                                                                                                                                                                                                                                                               \\ \hline

 \rowcolor{light-gray} 
  &   & \multicolumn{2}{c|}{ \begin{tabular}{c} Approach4 \\ Lognormal   \end{tabular}}     
	& \multicolumn{2}{c|}{  \begin{tabular}{c} Approach5\\ GEV \end{tabular}}                                                 
 	\\  \hline 
 
 \rowcolor{light-gray} 
\multirow{-2}{*}{\begin{tabular}{@{}c@{}}T \\ (sec)\end{tabular}} 
&   \multirow{-2}{*}{\begin{tabular}{@{}c@{}}$\upsilon (T)$  \\ Tbps$^{2}$\\  \end{tabular}}

&   \begin{tabular}[c]{@{}c@{}}  C4 \\ Mbps\end{tabular} &   \begin{tabular}[c]{@{}c@{}}Emp.\\ $\hat{ \varepsilon}$\end{tabular} &  
    \begin{tabular}[c]{@{}c@{}}C5 \\ Mbps\end{tabular} &  \begin{tabular}[c]{@{}c@{}}Emp.\\ $\hat{ \varepsilon}$\end{tabular} \\ \hline
0.01                                                                                       & 5.44                                                                                               & 48.186                                             & 0.0014                                             & 37.731                                             & 0.0076                                            \\ \hline
0.05                                                                                       & 2.56                                                                                               & 29.498                                             & 0.0052                                             & 27.632                                             & 0.0083                                            \\ \hline
0.1                                                                                        & 2.04                                                                                               & 26.732                                             & 0.0057                                             & 25.516                                             & 0.0083                                            \\ \hline
0.5                                                                                        & 1.16                                                                                               & 22.032                                             & 0.0072                                             & 21.428                                             & 0.0083                                            \\ \hline
1                                                                                          & 8.88                                                                                               & 21.110                                             & 0.0056                                             & 19.796                                             & 0.0096                                            \\ \hline
\end{tabular}
\end{table}

\begin{figure}[hpt]
 	\centerline{\includegraphics[scale=0.13]{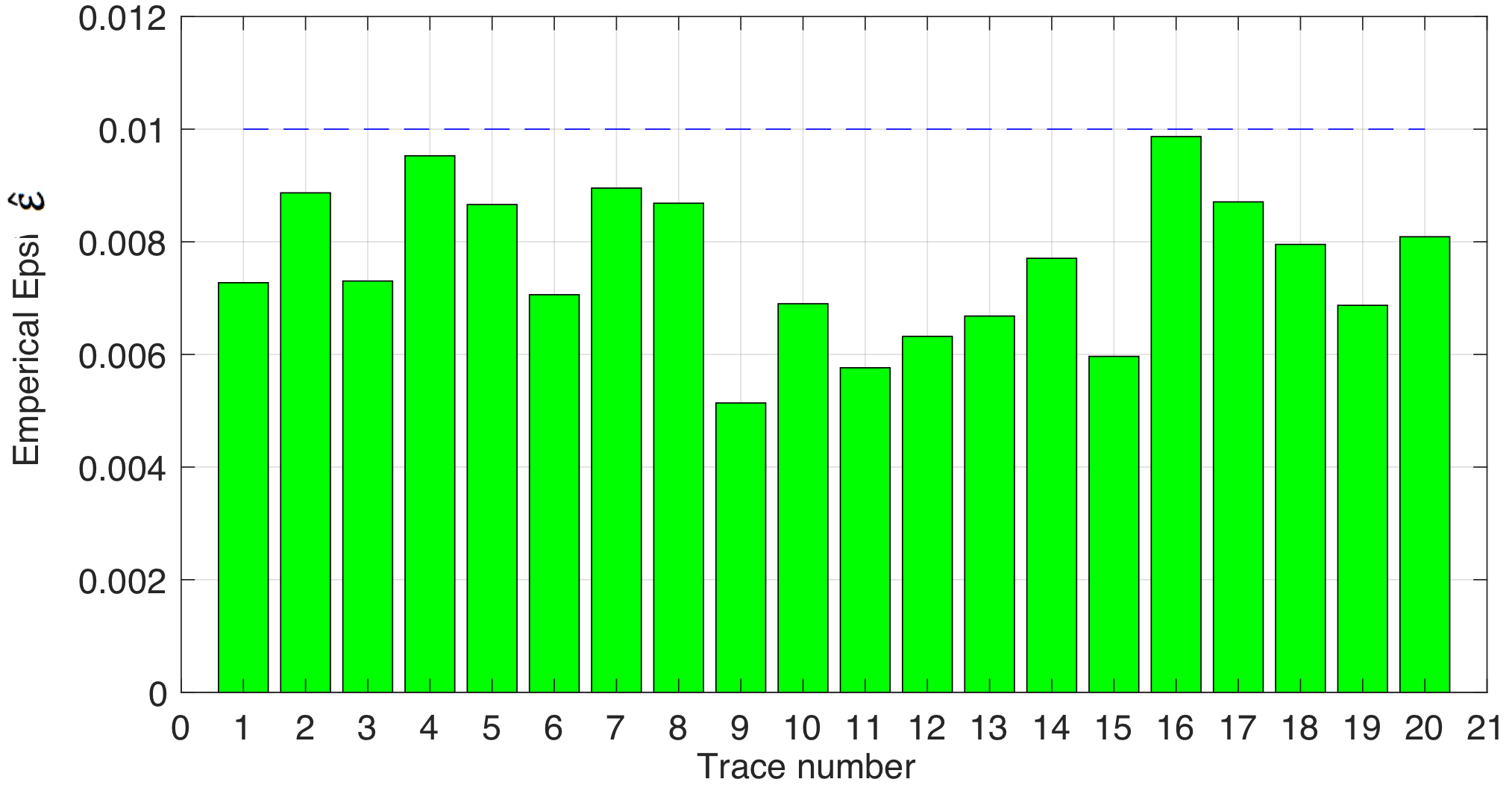}}
 	\caption{The empirical performance criterion $\hat{ \varepsilon}$ of 20 traces, when $T=0.01 sec$ and $\varepsilon  =0.01$ based on a GEV model}
 	\label{emp-app5}
 \end{figure}

 Fig. \ref{emp-app5} shows the results of applying approach 5 on 20 different traces captured at different locations. It is obvious that this approach provides a satisfied performance, as all the measured empirical performance criteria $\hat{ \varepsilon}$  values are less than 0.01. These results show that our objectives have been achieved.

 \section{Comparison between the five bandwidth provisioning approaches   }
 In this section we demonstrate the results of the five approaches using our GUI tool \cite{github}. The tool takes the captured traffic, the aggregation time T and the performance criterion $ \varepsilon$ as inputs. It calculates the link capacity of each approach (C1, C2, C3, C4 and C5). Besides, it gives the value of the empirical performance criterion $\hat{ \varepsilon}$ at every T value.   
 
\begin{figure}[htbp]
 	\centerline{\includegraphics[scale=0.24]{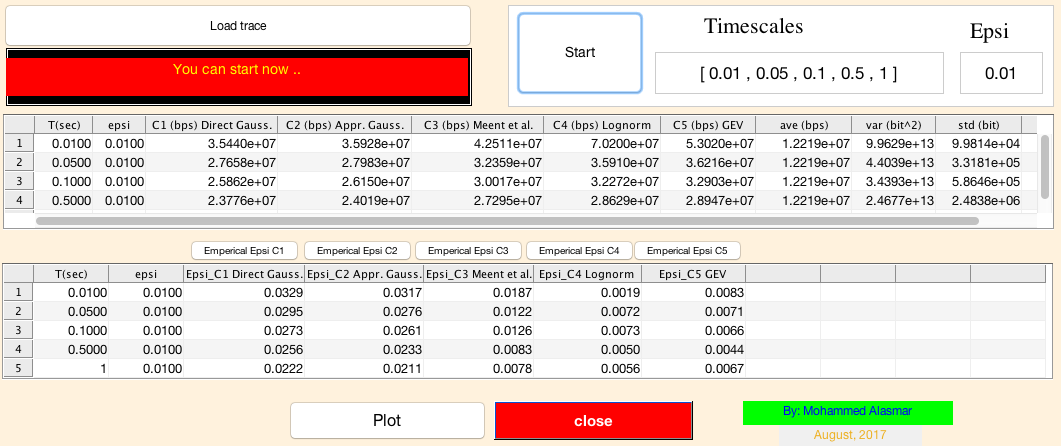}}
 	\caption{MATLAB GUI tool to perform the five bandwidth provisioning approaches}
 	\label{emploc5new}
 \end{figure}

In Fig.\ref{emploc5new}, the aggregation times have been passed as a vector (T=[0.01, 0.05, 0.1, 0.5 1])  and   $ \varepsilon = 0.01$. Trace 5 has been loaded to the tool and the measured capacities have been displayed. 
 
\begin{figure}[hpt]
	\centerline{\includegraphics[scale=0.34]{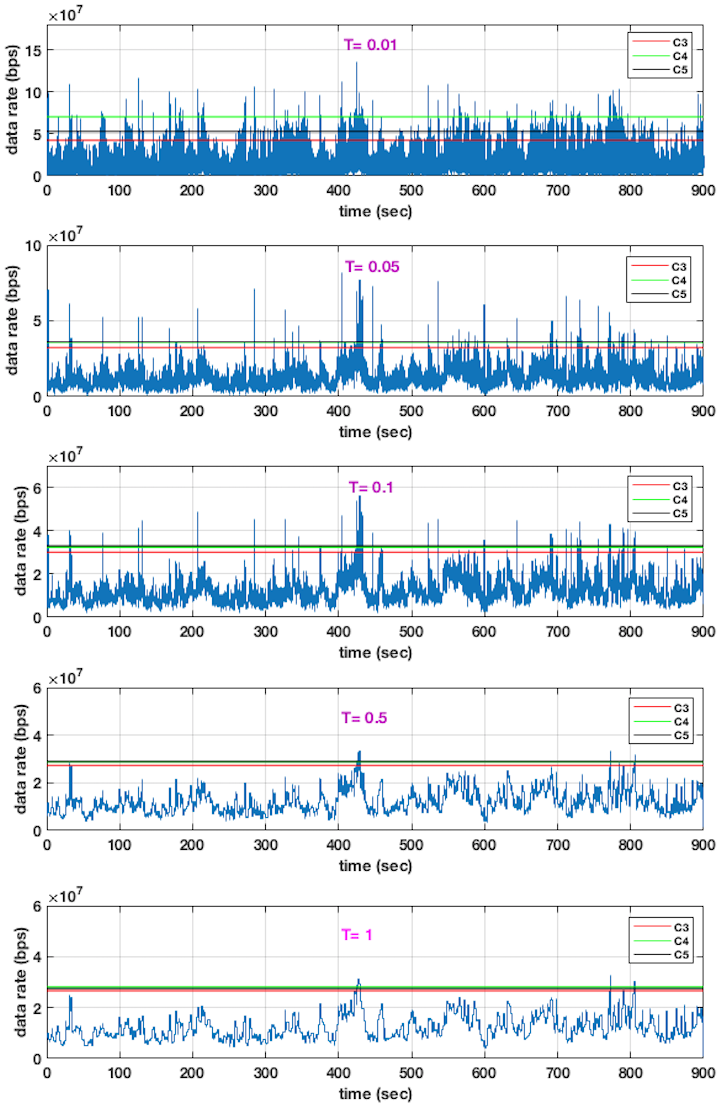}}
	\caption{The captured traffic A(T) and the estimated link capacities at different timescales}
	\label{AplotVSt}
\end{figure}

\begin{figure}[hpt]
	\centerline{\includegraphics[scale=0.25]{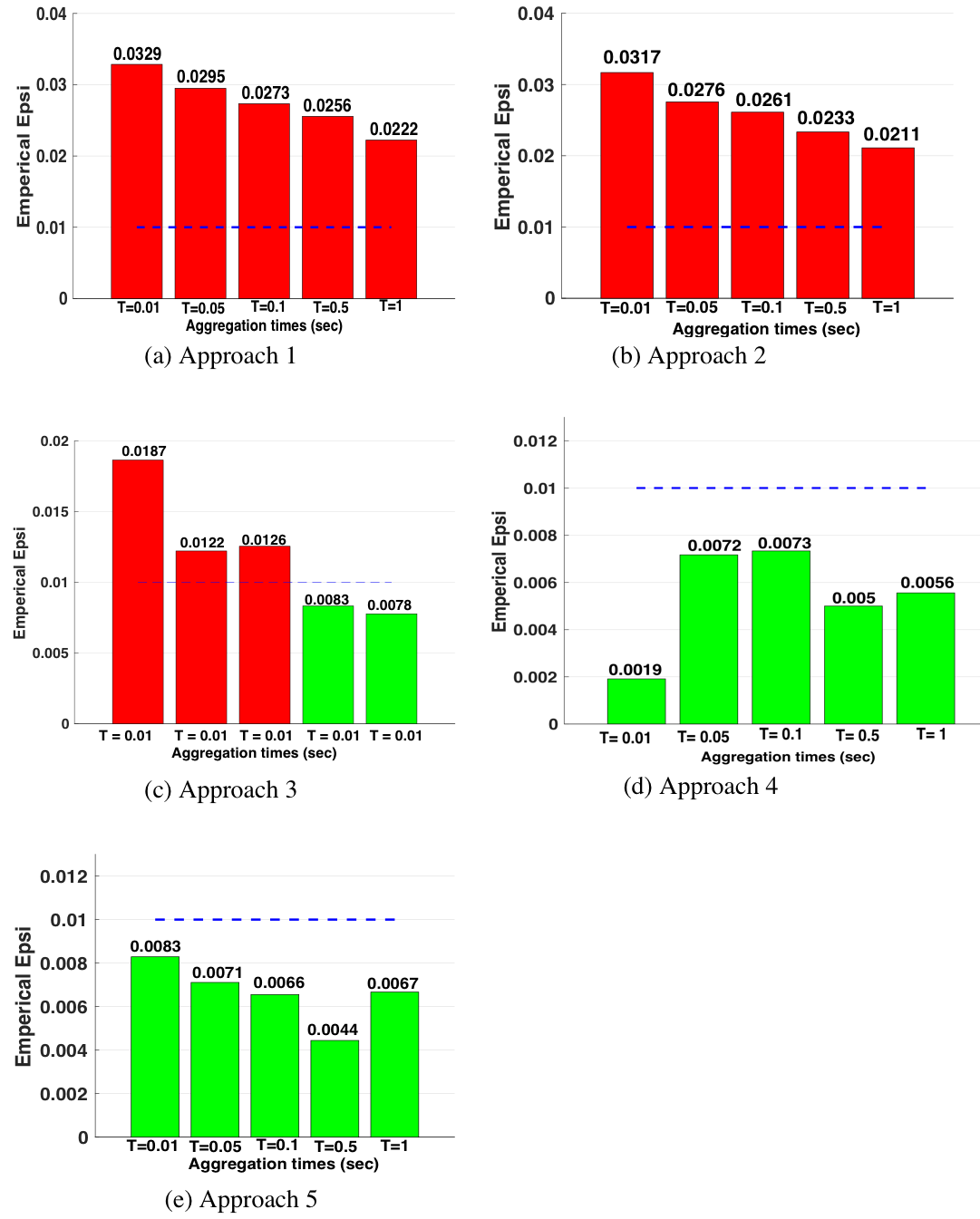}}
	\caption{The empirical performance criterion $\hat{ \varepsilon}$ of trace5, when $\varepsilon  =0.01$}
	\label{new-emp-tool}
\end{figure}

Furthermore, the tool can plot the captured traffic at different timescales, as illustrated in Fig.\ref{AplotVSt}. This figure shows that the traffic has larger data rates at small aggregation times. Moreover, the tool can display bar graphs that compare between the empirical performance criterion $\hat{ \varepsilon}$  and the target  performance   $\varepsilon $, as shown in Fig.\ref{new-emp-tool}. The red bars represent the failure of providing a satisfied performance(as $\hat{ \varepsilon} > 0.01 $), which is the case of approach 1 and approach 2 and mostly approach 3. In contrast, approach 4 and approach 5 show green bars, which means that the approaches are able to meet the SLA terms ($\hat{ \varepsilon} < 0.01 $).

 \newpage

\section{CONCLUSION}
  The principles behind the self-similarity phenomenon and how it affects the Internet traffic modeling is the most obvious finding to emerge from this research.  The statistical tests over a real Internet traffic shows that the heavy-tailed distributions such as lognormal and GEV are optimal in characterising the Internet traffic. Conversely, the light-tailed models as Gaussian distribution is not convergence with the tails.

This paper provides a generic methodology for link dimensioning.The successful implementation of an efficient bandwidth provisioning in an IP networks will contribute to helping
solve the limited bandwidth problem and prevent failures of links.  It was demonstrated that the capacity of Internet links can be accurately estimated using a simple mechanism, which requires a performance parameter that reflect the desired performance level and should be chosen by the network manager. Besides, it requires measuring the average link load and variance, which reflect the characteristics of the captured traffic.

The validation showed that our approaches(4 and 5) were able to determine the required link capacity accurately; our approach therefore clearly outperforms the simple rules of thumb that are usually relied on in practice.

\bibliographystyle{IEEEtran}
\bibliography{paperLatex}

\end{document}